\documentclass{aa}  
\usepackage{natbib}
\usepackage{graphicx}
\usepackage[dvipsnames]{xcolor}

\begin{document}

   \title{Investigating the impact of the dynamic solar wind on the propagation of a coronal mass ejection with two models and multi-spacecraft measurements}

   \author{T. Baratashvili \inst{1}, E. Davies \inst{2}, E. Weiler \inst{2,3}, B. Schmieder \inst{1,4,5}, P. Démoulin \inst{4}, S. Poedts \inst{1,6}
          }

   \institute{Department of Mathematics/Centre for mathematical Plasma Astrophysics, 
             KU Leuven, 3001 Leuven, Belgium\\
             \email{tinatin.baratashvili@kuleuven.be}
             \and
             Austrian Space Weather Office, GeoSphere Austria, Graz, Austria
             \and
             Institute of Physics, University of Graz, Universit\"atsplatz 5, 8010 Graz, Austria
             \and
             LIRA, Observatoire de Paris, Universit\'e PSL, CNRS, Sorbonne Universit\'e, Univ. Paris Diderot, Sorbonne Paris Cit\'e, 5 place Jules Janssen, 92195 Meudon, France
             \and LUNEX EMMESI COSPAR-PEX Eurospacehub, Kapteyn straat 1, Noordwijk 2201 BB, Netherlands
             \and
             Institute of Physics, University of Maria Curie-Sk{\l}odowska, Pl.\ Marii Curie-Sk{\l}odowskiej 1, 
             PL-20 031 Lublin, Poland
             }

   \date{Accepted: October 11, 2025}

  \abstract
   {Coronal mass ejections (CMEs) are the main drivers of disturbances in the solar heliosphere because they propagate and interact with the magnetic field of the solar wind.
   It is crucial to investigate the evolution of CMEs and their deformation for understanding the interaction between the solar wind and CMEs.
   }
   {We quantify the effect of the dynamic solar wind on the propagation of a CME in the heliosphere with a hydrodynamic plasma cloud-cone model and a linear force-free spheromak model at various locations in the heliosphere. 
   }
   {We chose a CME event that launched on SOL2021-09-23T04:39:45 and was observed by multiple spacecraft, namely BepiColombo, Parker Solar Probe, Solar Orbiter, Stereo A and ACE. The solar wind was modelled in the steady and dynamic regimes in the Icarus model. The CME parameters were approximated for the selected event, and two CME models (spheromak and cone) were launched from the inner heliosphere boundary. The obtained synthetic in situ measurements were compared to the observed in situ measurements at all spacecraft. 
   }
   {The internal magnetic field of the flux rope was better reconstructed by the spheromak model than by the cone CME model. 
   The cone CME model maintained a nearly constant longitudinal angular extension while somewhat contracting in the radial direction. In contrast, the spheromak model contracted in the longitudinal direction while expanding in the radial direction. 
   }
   {The CME sheath and magnetic cloud signatures were better reproduced at the four spacecraft clustered around the CME nose by the spheromak CME model. The dynamic solar wind caused a greater deceleration of the modelled CME than the steady-state solar wind solution. Because the background was homogeneous, the modelled CME properties were only mildly affected by the solar wind regime, however. 
   }

   \keywords{Magnetohydrodynamics (MHD); Methods: numerical; Methods: observational; Sun: coronal mass ejections (CMEs); Sun: heliosphere;  }
\titlerunning{The impact of dynamic solar wind on CME propagation}
\authorrunning{Baratashvili et. al}
\maketitle

\section{Introduction} \label{sec:Introduction}
Coronal mass ejections (CMEs) are violent eruptions originating from the solar corona. CMEs are considered the primary drivers of space weather \citep{Gopalswamy2017}. Space weather is a branch of physics that focuses on the disturbances in the solar heliosphere with a particular emphasis on the near-Earth environment. During {a} CME eruption, an enormous amount of plasma is released into the solar atmosphere, carrying up to $10^{16}\;$g of material \citep{Webb2012}, and propagating with speeds ranging from $300\;$km s$^{-1}$ to $1000\;$km s$^{-1}$, according to observations with the Solar and Heliospheric Observatory/Large Angle and Spectrometric Coronagraph \citep[SOHO/LASCO,][]{Domingo1995,Brueckner1995}. 
The interior of the CME is not homogeneous. It often has an internal magnetic field with a magnetic flux rope in which twisted magnetic field lines are assumed to be helically wound around a central axis \citep{Webb2012, Cane2003, Vourlidas2013}. A lower proton temperature than expected in the solar wind, with a similar velocity, is typically present, along with many other possible properties (e.g., \cite{Zurbuchen2006}). This cold plasma with a rotating magnetic field is called a magnetic cloud \citep{Burlaga1981, Klein1982}. 

Coronal mass ejections are usually associated with solar eruptive flares. The same active region can release multiple CMEs over a short period \citep[e.g.][]{Demoulin2002,Green2002}. A higher frequency of CMEs usually occurs during solar maximum, when the Sun is more active and stronger magnetic fields are present in the solar photosphere. The
CMEs propagate through the solar wind, which is a continuous outflow of magnetised plasma from the solar corona. Its structure is highly complex and dynamic, characterised by a bi-modal velocity distribution \citep{McComas1998}. 
Since the solar wind is also magnetised, the CMEs interact with it when they propagate, sometimes resulting in deformation, erosion or deflection \citep{Gopalswamy2017}. When multiple CMEs are travelling from the Sun to Earth, they can interact with each other \citep[e.g.][]{Dasso2009}, which increases their complexity \citep{Scolini2022, Davies2022}. \cite{lee2013} showed that a more accurate solar wind simulation yields better predictions of the CME arrival and strength at 1\;au.

Space-weather forecasting focuses on predicting the arrival of the CME at Earth and its potential impact. The primary challenge in space-weather forecasting for models is the efficiency and robustness of the models themselves. Recently, a few space-weather modelling toolchains have been established. Icarus \citep{Verbeke2022, Baratashvili2025} is a 3D time-dependent magnetohydrodynamic (MHD) heliosphere model developed at the Centre for Mathematical Plasma-Astrophysics (CmPA; KU Leuven). It can be driven by a semi-empirical Wang-Sheeley-Arge (WSA) coronal model \citep{arge2003}, or with a full 3D MHD model suitable for space-weather modelling, such as COolFuid COroNa UnsTructured (COCONUT) \citep{Baratashvili2024C}. Icarus is implemented within \texttt{MPI-AMRVAC} \citep{Keppens2023} and supports advanced techniques, such as adaptive mesh refinement (AMR) and radial grid stretching, to achieve fast and efficient simulations. 

The standard approach for space-weather tools is to drive simulations with steady boundary conditions. This involves selecting a particular magnetogram, computing the solar corona for the given magnetogram, and providing the plasma conditions at 0.1~au for the heliosphere model. This often leads to an outdated solar wind in the simulation, that omits new information observed in the solar atmosphere and limits the simulations to the Earth's orbit. 
Recently, Icarus was upgraded to support time-dependent boundary driving. This means that the inner boundary conditions are continuously updated after the simulation starts to mimic the realistic solar wind. In this way, the solar wind can be dynamic, and the {outer boundary of the} heliosphere {model} can be extended farther with continuously updated conditions in the global heliosphere. The most recent developments in Icarus enable us to investigate the impact of the dynamic solar wind on the propagation of CMEs. 

It is currently not possible to observe global CME structures because our observation capabilities are limited. Their projection on the plane of sky can be observed with coronagraphs such as SOHO/LASCO-C2 and C3 \citep{Brueckner1995}, or with the coronagraphs mounted on two spacecraft of the Solar Terrestrial Relations Observatory \citep[STEREO,][]{Kaiser2005}. The coronagraphs provide an estimate of the geometrical characteristics of the CME, including its plane of sky, projected direction of propagation and speed, but they lack information about the internal magnetic structure of the CME. 

The internal structure of a CME can be studied in detail through in situ measurements. This approach only samples over a time-dependent spacecraft trajectory, representing a single 1D slice through a larger 3D structure, however. Even in these cases, it is often difficult to conclude about the global evolution and deformation of the CME throughout its journey from the Sun to Earth. \cite{Bourlaga1982} suggested that considering multi-point observations of the CME might decrease the uncertainty associated with the propagation of the global structures. There are currently no multi-point missions for studying CMEs over useful separations in operation, and scientists therefore rely on luck to detect CME events that have been observed by multiple spacecraft. Various catalogues and studies focused on multi-spacecraft events, making them more accessible to scientists \citep{Grison2018, Salman2020, Davies2020,  Davies2021, Palmerio2021, Moestl2022, Davies2022, Baratashvili2024B}.

We investigated the propagation of CMEs in configuration of the steady and dynamic solar wind. We selected the cone and a magnetised CME model to evaluate the effect on the internal CME magnetic field better. We selected the CME that was observed by multi-spacecraft previously reported in the Helio4Cast Lineup catalogue \citep{Moestl2022}, which lists multi-point CMEs, with each CME linked to the HELIO4CAST ICMECAT \citep{Moestl2020}. Furthermore, the event was studied by \cite{Palmerio2025}. The CME erupted on September 23, 2021, and was observed by five spacecraft at various radial distances, including BepiColombo \citep[Bepi,][]{Benkhoff2021, heyner2021bepicolombo}, Solar Orbiter \citep{Muller2020}, Parker Solar Probe \citep[PSP,][]{Fox2016}, STEREO-A \citep[STA,][]{Kaiser2005}, and Advanced Composition Explorer \citep[ACE,][]{Stone1998}. The distribution of the spacecraft enabled us to sample the CME at various radial distances and angular separations. The synthetic in situ data were compared to the observed measurements, and the evolution of the hydrodynamic and magnetised CME 3D structures was investigated.

The paper is organised as follows: Section \ref{sec:Observational_Evidence} summarises the observational evidence in the solar atmosphere and in situ. The numerical setup of Icarus is described in Section~\ref{sec:Numerical}. The results of the simulations and their comparison with the in situ measurements are given in Section~\ref{sec:Results}, and the conclusions are summarised in Section~\ref{sec:Conclusions}.

\section{Observational evidence}  \label{sec:Observational_Evidence}
\begin{figure*}[hpt!]
\centering
    \includegraphics[width=0.32\textwidth]{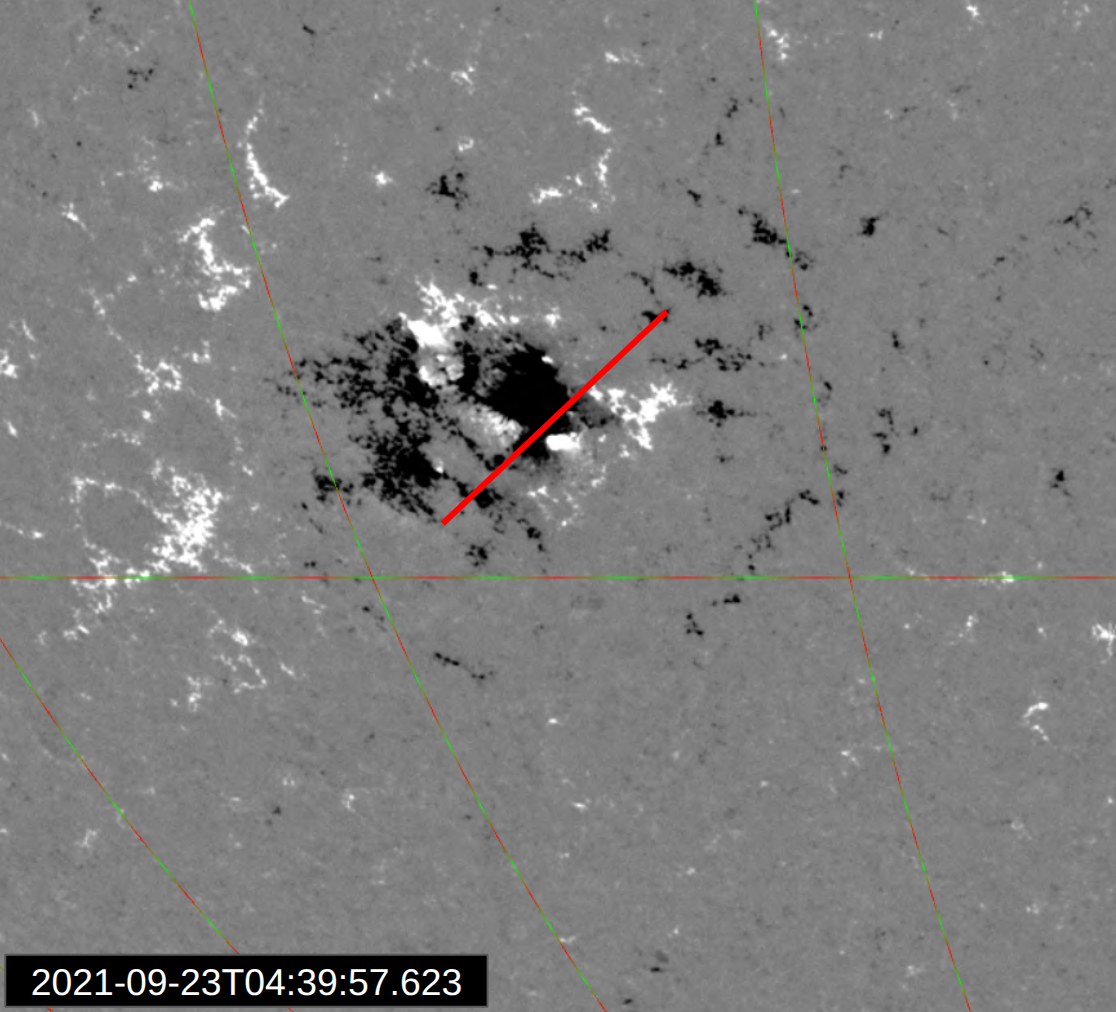}
    \hfill
    \includegraphics[width=0.32\textwidth]{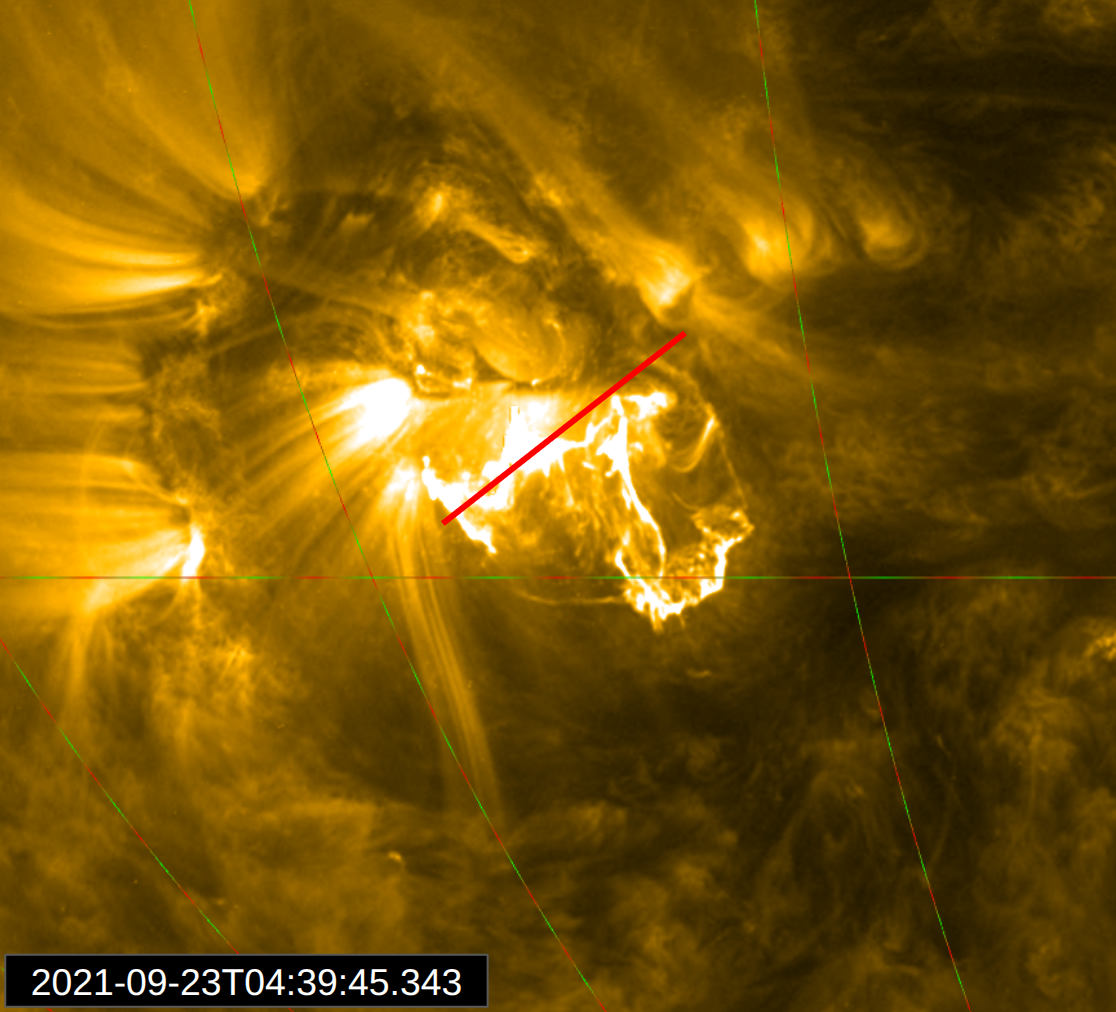}
    \hfill
    \includegraphics[width=0.32\textwidth]{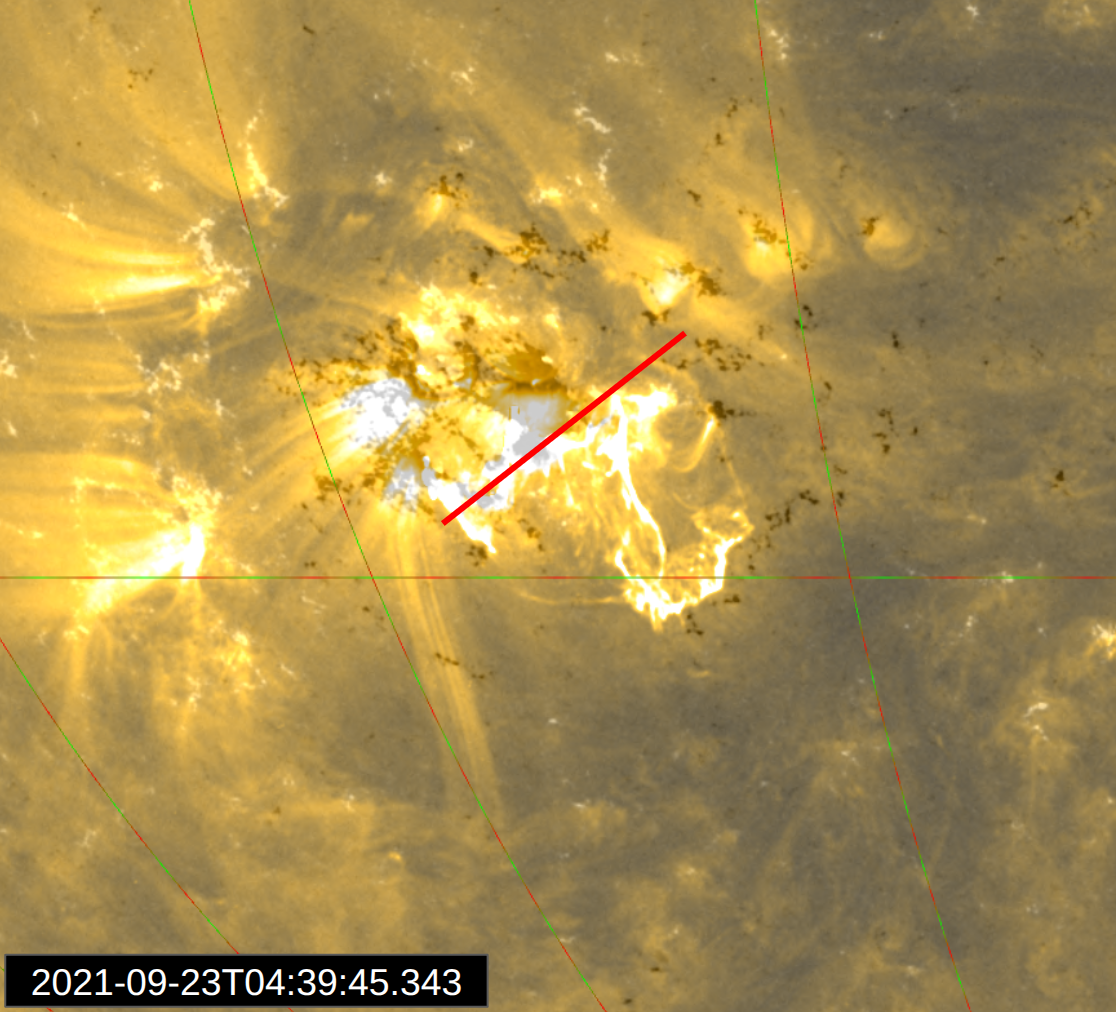}
  \caption{Configuration of the solar atmosphere during the CME eruption for SOL2021-09-23T04:39:45. The left panel represents the HMI magnetogram, which is saturated between $\pm 1000\;$G. The middle panel represents the AIA 171\AA\ image, and the right panel shows the AIA image overlaid on the magnetogram. The red line shows the approximate polarity-inversion line. The figures were created with JHelioviewer. } \label{fig:surface_AIA_HMI}
\end{figure*}

\begin{table*}[ht!]
\caption{Observed times of the shock, leading and trailing edges of the magnetic ejecta at multiple spacecraft.}   
\label{table:cme_arrival_times}   
\centering            
\begin{tabular}{l c c c c c c }         
\hline
& Shock & ME LE & ME TE & Sheath [h] & ME [h] & V$_{\text{B}}$ [km s$^{-1}$] \\ 
\hline\hline
  Bepi & 2021-09-25T01:46:00 & 2021-09-25T09:35:00 & 2021-09-25T18:50:00 & 7.8 & 9.25 &401.9 \\
  Solar Orbiter & 2021-09-25T18:25:00 & 2021-09-26T04:37:00 & 2021-09-27T00:00:00 & 10.2 & 19.4 & 407.9\\
  PSP & 2021-09-26T08:50:00 & 2021-09-26T23:04:00 & 2021-09-27T12:40:00 & 14.2 & 13.6 & 423.5 \\
  STA & 2021-09-27T01:51:00 & 2021-09-27T15:35:00 & 2021-09-28T04:31:00 & 13.7 & 12.9 & 426.4 \\
  ACE & 2021-09-27T04:16:00 & 2021-09-27T04:16:00 & {2021-09-28T09:25:00} & 29.2 & N/A & 428.7\\
\hline                                  
\end{tabular}
\tablefoot{The first column lists the spacecraft at which the CME was observed in situ. The second to the fourth columns show the arrival times in (UT) of the shock, ME LE, and ME TE at the observing spacecraft. The last three columns show the duration of the sheath and magnetic cloud in hours and the ballistic propagation speed in km~s$^{-1}$. The ballistic speed is calculated with V$_B$ = (spacecraft location-1 R$_\odot$)/(Shock-eruption time).}
\end{table*}

The CME was selected from the Helio4Cast Lineup catalogue \citep{Moestl2022} and has previously been studied by \citet{Palmerio2025}. The event is an excellent candidate for validating the model performance extensively because multi-point observations over large radial and angular extents are available.

\subsection{Solar atmosphere} 
The CME we chose to study erupted from the solar corona on September 23, 2021, at 04:40~UT. It originated from NOAA active region (AR) 12871, following the M2.8 flare that started at 04:35~UT, peaked at 04:42~UT, and ceased at 04:50~UT. The flare was located at S30E22, far from the equatorial plane and east of the Sun-Earth line. The active region was very complex, and it was challenging to determine the exact magnetic field configuration of the erupting structure. 

Figure~\ref{fig:surface_AIA_HMI} shows the source region of the CME. The left panel shows the Helioseismic and Magnetic Imager {(HMI; \citealt{Scherrer2012})} magnetogram at SOL2021-09-23T04:39:57 on board the Solar and Heliospheric Observatory {(SDO; \citealt{Pesnell2012})}. 
The figure was taken from JHelioviewer \citep{Muller2017}, and the values are therefore saturated to $\pm1000$ G.
The middle panel shows the image taken by the Atmospheric Imaging Assembly {(AIA; \citealt{Lemen2012})} at SOL2021-09-23T04:39:45 on board SDO through the 171 \AA\ filter. 
The last panel shows the AIA image overlaid on the magnetogram. The same red line is plotted on all figures as a crude approximation of the polarity inversion line (PIL) for the region from which the CME of interest erupted. 
This linear and straightforward approximation suggests an inclination of $\sim43^\circ$, which we used to calculate the tilt of the flux rope in the simulation. 
The CME had a positive magnetic helicity \citep[using helicity sign proxies as summarised in Figure 1 of][]{DemoulinPariat2009}, resulting in a right-hand magnetic field configuration. 

\begin{figure}[hpt!]
\centering    \includegraphics[width=0.49\textwidth]{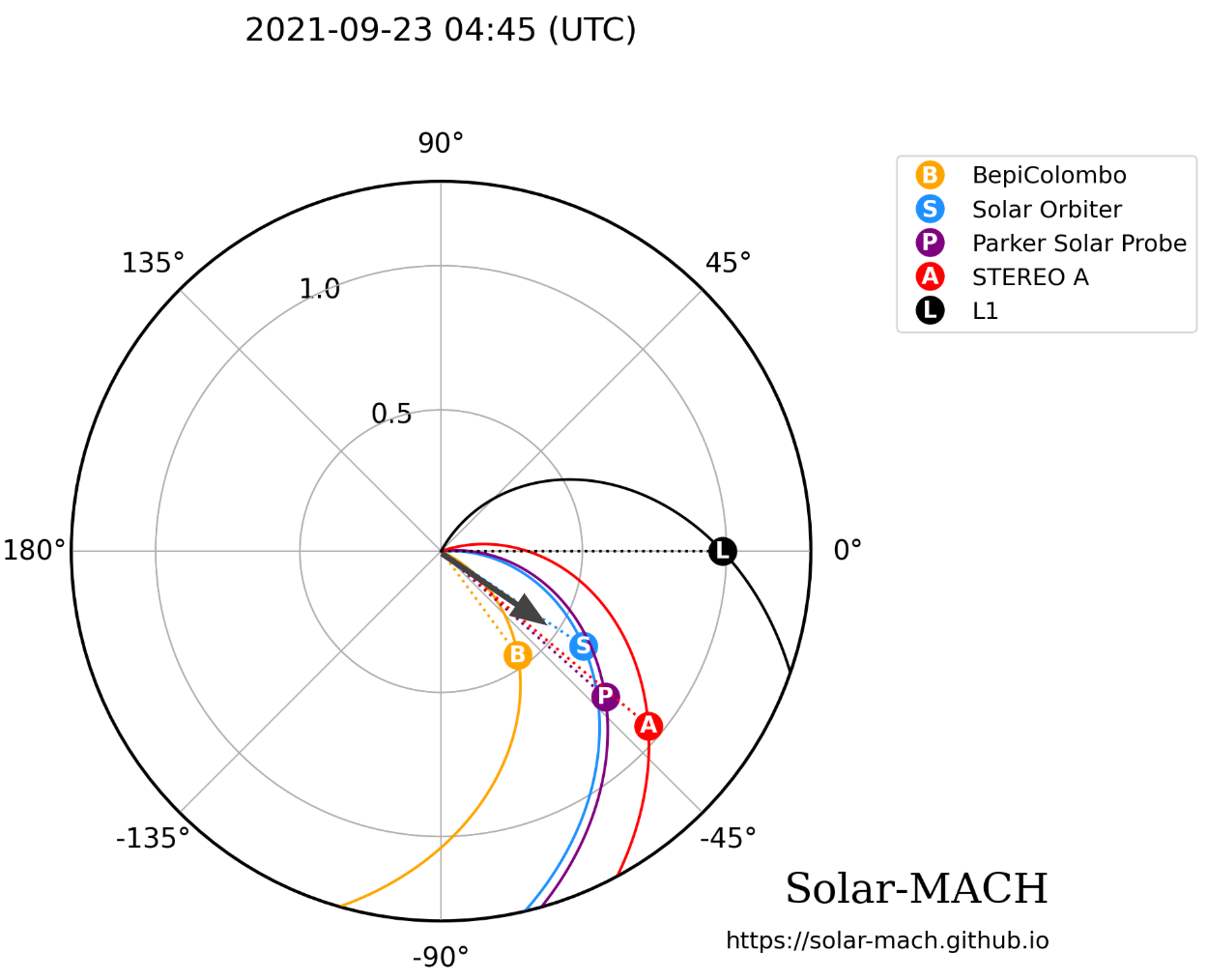}
  \caption{Location of the spacecraft in Stonyhurst coordinates during the eruption of the CME in a top-down view on the ecliptic plane. The theoretical Parker spiral passing at each spacecraft was added. The arrow denotes the propagation direction of the CME. }
  \label{fig:insitu_satellties}
\end{figure}

\begin{table}[ht!]
\caption{Spacecraft location at the shock times in Table~\ref{table:cme_arrival_times} in Stonyhurst coordinates.}   
\label{table:spacecaft_locations}   
\centering            
\begin{tabular}{l c c c}         
\hline
& R [au] & $\theta$ [$^\circ$] & $\phi$ [$^\circ$]  \\ 
\hline\hline
  BepiColombo & 0.44 & -49.3 & -0.1  \\
  Solar Orbiter & 0.61 & -30.1 & 1.9  \\
  Parker Solar Probe & 0.78 & -42.8 & 3.5  \\
  STEREO A & 0.96 & -49.8 & 6.9  \\
  ACE  & 0.99 & 0.0 & 7.0  \\
\hline                                  
\end{tabular}
\tablefoot{$\theta,\phi$ denote longitude and latitude, respectively. }
\end{table}

\label{sec:numerical_setup}
\begin{figure*}[hpt!]
\centering
    \includegraphics[width=\textwidth]{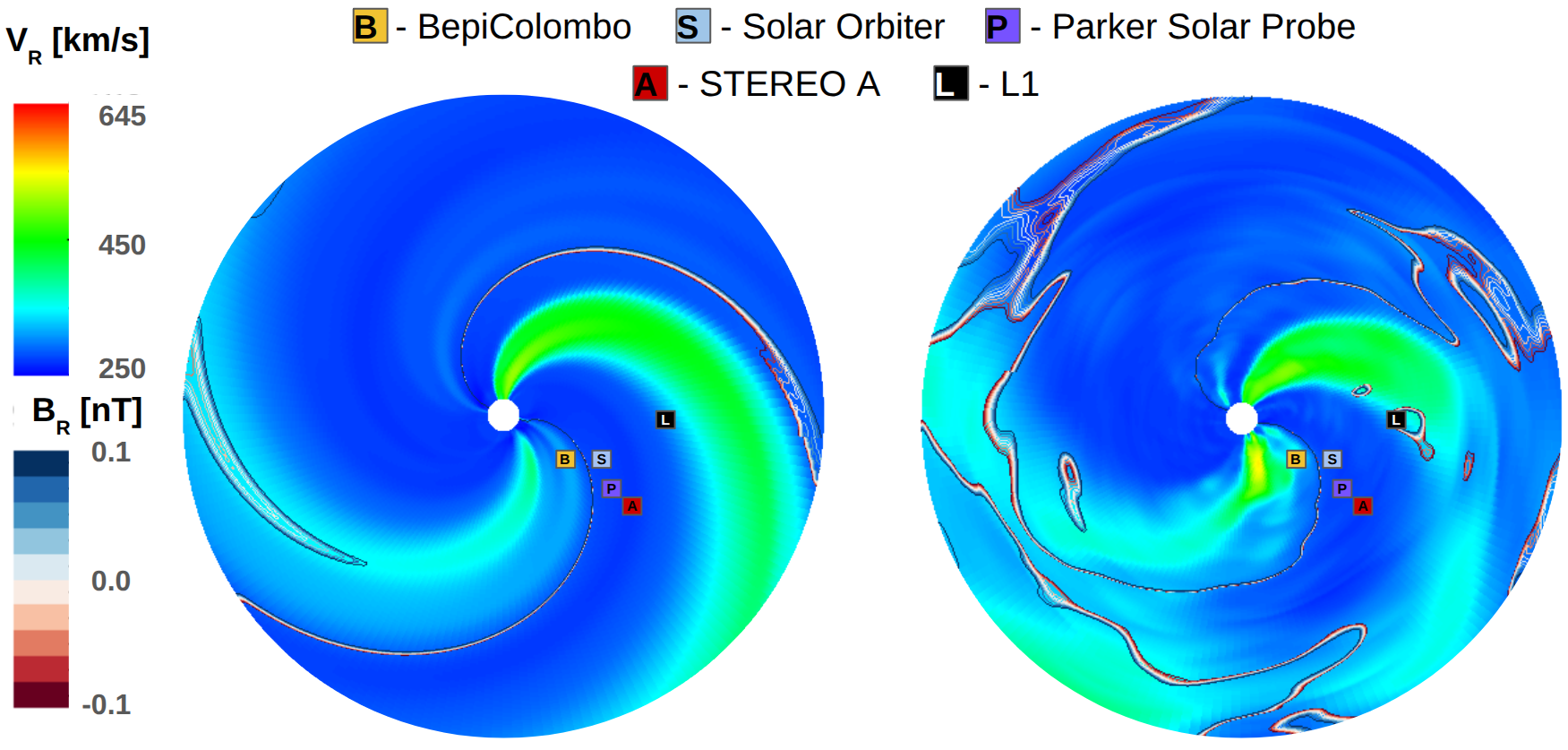}
  \caption{Steady (left) and dynamic (right) solar wind configurations before the CME injection at 0.1~au. The radial velocity is plotted with colour levels in the solar equatorial plane, along with the contours of the magnetic field component, $B_r$, to identify the heliospheric current sheet (HCS). The spacecraft are plotted on top of the solar wind.
  } \label{fig:steady_vs_dynamic_wind}
\end{figure*}

\subsection{In situ Observations}
The CME was reported in the Helio4Cast LineupCAT (V3.0 \footnote{\url{https://helioforecast.space/lineups}}, ID: MULTI\_2021-09-23\_01) and was expected to be observed by five spacecraft: BepiColombo, Solar Orbiter, Parker Solar Probe, STEREO-A, and ACE (at L1). The configuration of the spacecraft is given in Figure~\ref{fig:insitu_satellties} in Stonyhurst coordinates at the time of the eruption, and the arrow shows the estimated propagation direction of the CME. The figure was generated with Solar-MACH \citep{Gieseler2022}. 

Table~\ref{table:cme_arrival_times} summarises the arrival times of the CME shock, the leading (ME LE) and trailing edges of the magnetic ejecta (ME TE), and the durations of the sheath and the magnetic cloud at each spacecraft. The CME shock arrived at BepiColombo, the spacecraft nearest to the Sun, almost two days after the eruption, and it reached the Earth four days after the eruption. We were unable to identify the magnetic cloud at L1 (ACE) and only detected the signatures of the sheath, however. 
There was an apparent minor acceleration of the CME as it travelled through the inner heliosphere (see the ballistic speed calculations in the final column of Table~\ref{table:cme_arrival_times}). The separation in the  two $(\theta,\phi)$ directions was notable, however, because it indicated that different portions of the CMEs were crossed by other spacecraft, as shown in Table~\ref{table:spacecaft_locations}. 
The CME structure was observed from 0.44~au to 1~au, spanning longitudinal angles from $0^\circ$ to $-49.3^\circ$ and latitudinal angles from $-0.1^\circ$ to $7.0^\circ$. This means that each spacecraft interacted with a different region inside the CME.

\section{Numerical setup}  \label{sec:Numerical}
\subsection{Icarus software}

Icarus \citep{Baratashvili2025,Verbeke2022,Baratashvili2022} is a 3D MHD heliosphere modelling tool developed in the framework of \texttt{MPI-AMRVAC} \citep{Keppens2023}. Its domain extends from 21.5 R$_\odot$ (0.1~au) to 432.5 R$_\odot$ (2~au) in the radial direction, spans $360^\circ$ in the longitudinal direction and extends to $\pm60^\circ$ from the equatorial plane. The MHD equations were solved in the co-rotating frame by adding the centrifugal and Coriolis forces to the equations  \begin{align}\label{eq:mhd}
    \frac{\partial \rho}{\partial t}+ \nabla \cdot (\rho \mathbf{v}) & =0,\\
    \frac{\partial (\rho\mathbf{v})}{\partial t} + \nabla \cdot \bigg(\rho\mathbf{v}\mathbf{v}+p_{tot}\mathbf{I}-\mathbf{B}\mathbf{B} \bigg) - \rho \mathbf{g} &= \mathbf{F}, \label{eq:icarus_momentum}\\
    \frac{\partial e}{\partial t} + \nabla \cdot \bigg( e\mathbf{v} + p_{tot}\mathbf{v} - \mathbf{B}(\mathbf{B}\cdot \mathbf{v}) \bigg) - \rho\mathbf{v} \cdot \mathbf{g}  & = \mathbf{v} \cdot \mathbf{F} , \label{eq:icarus_energy} \\
     \frac{\partial \mathbf{B}}{\partial t} + \nabla\cdot \bigg(\mathbf{v}\mathbf{B} - \mathbf{B}\mathbf{v}\bigg)&=0,\label{eq:faraday} \\
    \nabla \cdot \mathbf{B} &=0,
\end{align}
where
\begin{equation}
\quad p_{tot} = (\gamma-1)\bigg(e-\rho \frac{\mathbf{v}^2}{2}-\frac{\mathbf{B}^2}{2}\bigg) + \frac{\mathbf{B}^2}{2},
\end{equation}
and $\rho$, $\mathbf{v}$, $p_{tot}$, $\mathbf{B}$, and $e$
are the mass density, the velocity vector field, the total pressure (the sum of the thermal and magnetic pressure) of the plasma, the magnetic field vector, and the total energy density, respectively. We refer to \cite{Porth2014} for more details about the implementation in the \texttt{MPI-AMRVAC} framework. 

We considered the magnetic permeability $\mu_0$ to be one in non-dimensional units. The gravitational acceleration $\mathbf{g}$ is given by $({GM_{\odot}}/{r^2}\mathbf{e_r})$, with $G$ the gravitational constant and $M_{\odot}$ the mass of the Sun. The polytropic index $\gamma$ is 1.5, similar to \cite{Pomoell2018} and \cite{Odstrcil2004}. This reduced index, compared to the classical adiabatic index $\gamma=5/3$, models additional heating in the simplest way and yields solar wind acceleration throughout the inner heliosphere \citep[see, for example,][]{Pomoell2012}. Finally, the source term $\mathbf{F}$ is given by $\rho (\mathbf{\Omega} \times \mathbf{r}) \times \mathbf{\Omega} + 2 \rho (\mathbf{v}\times\mathbf{\Omega})$, corresponding to the centrifugal and Coriolis forces, respectively. Here, $\Omega$ (the magnitude of the vector $\mathbf{\Omega}$) is the rotation rate of the Sun at the equator ($2.97\cdot 10^{-6}\;$rad/s). The standard uniform grid was used for the simulations with the resolution in [R,$\theta$,$\phi$] = [0.685\;R$_\odot$, $1.875^\circ, 1.875^\circ$].

The equations were solved with a shock-capturing second-order (in time and space) total vanishing diminishing Lax-Friedrichs (TVDLF) scheme \citep{TothOdstrcil1996} in combination with a Vanleer limiter \citep{vanleer1974}. A parabolic cleaning was applied to minimise $\nabla \cdot \mathbf{B}$ errors, introduced in Eq.~(8) of \cite{Keppens2003}, which introduced diffusive terms in the energy and induction equations. 

The latest release of the Icarus model allows the dynamic inner boundary driving of the simulations. \cite{Baratashvili2025} introduced the procedure for generating the dynamic solar wind simulations. 
The outputs of the coronal model at 21.5 R$_\odot$ were combined into a datacube that was used as an inner boundary condition for Icarus. The cadence of the updated inner boundary conditions can be arbitrary; it was constrained by the available observations. The information was updated continuously, without relaxing the initial snapshot. Thus, the first eight days of the simulation time were taken as the so-called relaxation phase, during which the solar wind outflow was established. The CMEs were injected from the inner heliosphere boundary on top of the dynamic solar wind. 

\subsection{Steady and dynamic solar wind}
The WSA model \citep{arge2003} was used to generate the inner boundary conditions for the solar wind in the Icarus heliosphere. The boundary was prepared in two regimes: steady and dynamic. For the steady boundary driving, the Global Oscillation Network Group \citep[GONG,][]{harvey1996} magnetogram observed on September 23, 2021 04:04 (UT) was used. 

For the dynamic solar wind simulations, the GONG magnetograms between September 10, 2021 00:04 and October 10, 2021 00:14:00 (UT) were obtained with an average cadence of four hours. The WSA coronal models were computed individually for all 181 magnetograms, and the generated input boundary files were combined into the datacube that Icarus reads. The simulation was run until October 17, 2021 06:14 to ensure that the last updated magnetograms also had time to propagate to the outer boundary.

Figure~\ref{fig:steady_vs_dynamic_wind} shows the steady (left) and dynamic (right) solar wind in the equatorial plane. The snapshot of Figure~\ref{fig:steady_vs_dynamic_wind} shows the moment before the CME eruption to assess the impact of the dynamic boundary driving on the heliosphere. The high-speed stream profiles are notably disturbed by the dynamic boundary, with a smaller effect in the more homogeneous low-speed streams. All of the spacecraft are located in the low-speed stream; only BepiColombo and L1 are positioned near the interaction region between the high- and low-speed streams in the solar wind, but the impact on the dynamic HCS is not prominent at the radial distance of BepiColombo. 

\begin{figure}[ht!]
\centering
    \includegraphics[width=0.5\textwidth]{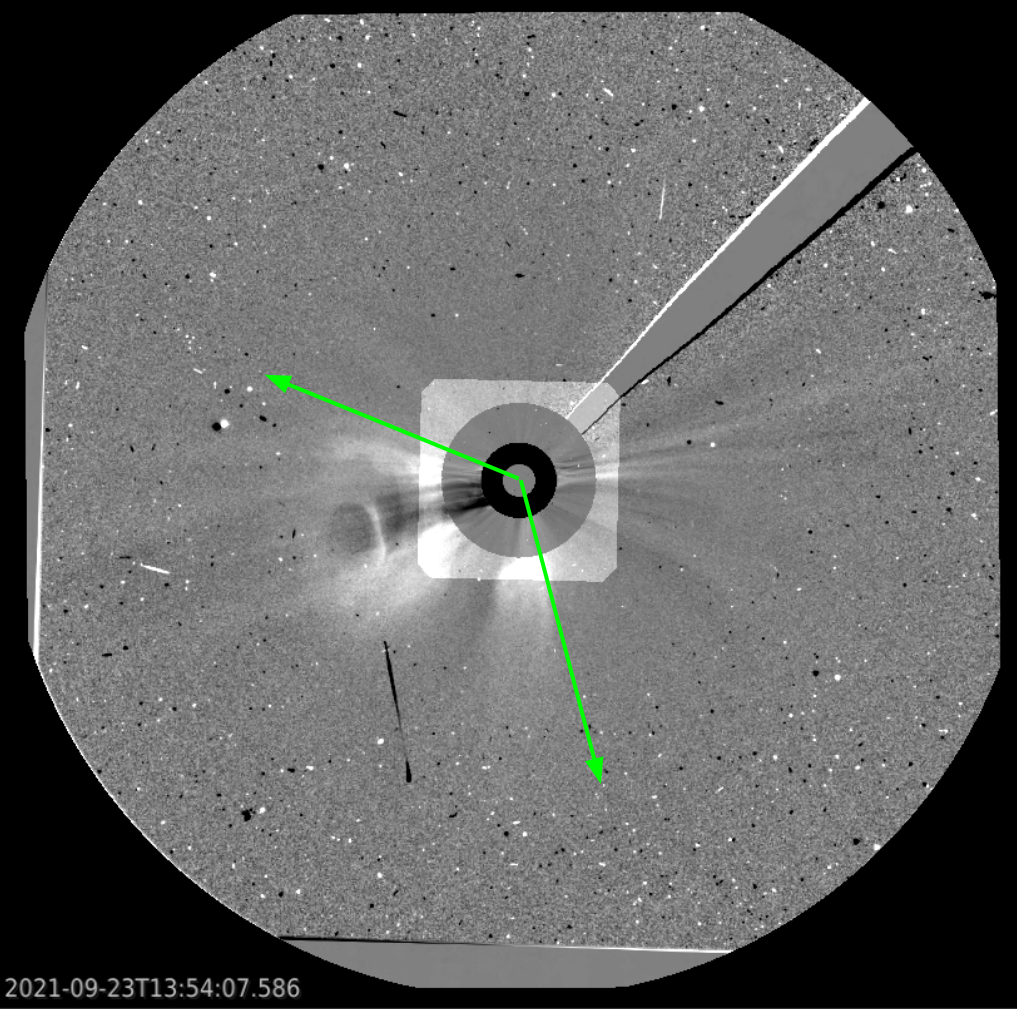}
  \caption{Base difference of the LASCO C2 and C3 images. The HMI magnetogram is plotted on the solar disk. Green arrows denote the CME extent.
  } \label{fig:LASCO_C2_C3}
\end{figure}

\begin{table}[ht!]   
\centering  
\caption{Input parameters for the cone and spheromak CME models.}  
\begin{tabular}{ccc}         
\hline 
 & Cone & Spheromak \\ 
\hline  \hline        
    t$_{\rm CME}$ & 2021-09-23T16:44:00 &  2021-09-23T16:44:00  \\ 
    $\theta_{\rm CME}$ & -43$^\circ$ & -43$^\circ$ \\ 
    $\phi_{\rm CME}$ & -6$^\circ$ & -6$^\circ$ \\
    r$_{\rm CME}$ & 20 R$_\odot$ & 20 R$_\odot$\\ 
    v$_{\rm CME}$ & 550 km s$^{-1}$ & 420 km s$^{-1}$\\ 
    $\rho_{\rm CME}$ & 1.5$\times10^{-18}$ kg m$^{-3}$ & 1.5$\times10^{-18}$ kg m$^{-3}$ \\
    T$_{\rm CME}$ &  $10^5$ K &$10^5$ K\\ 
    $\tau_{\rm CME}$ & N/A & 70$^\circ$ \\ 
    H$_{\rm CME}$ & N/A &+1\\ 
    F$_{\rm CME}$ & N/A & 0.2 $\times 10^{14}$ Wb\\
\hline                           
\end{tabular}
\label{table:CME_parameters}
\end{table}

\begin{figure*}[hbt!]
\centering
    \includegraphics[width=0.8\textwidth]{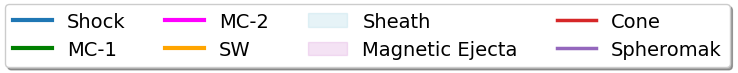}
    \hfill
    \includegraphics[width=0.49\textwidth]{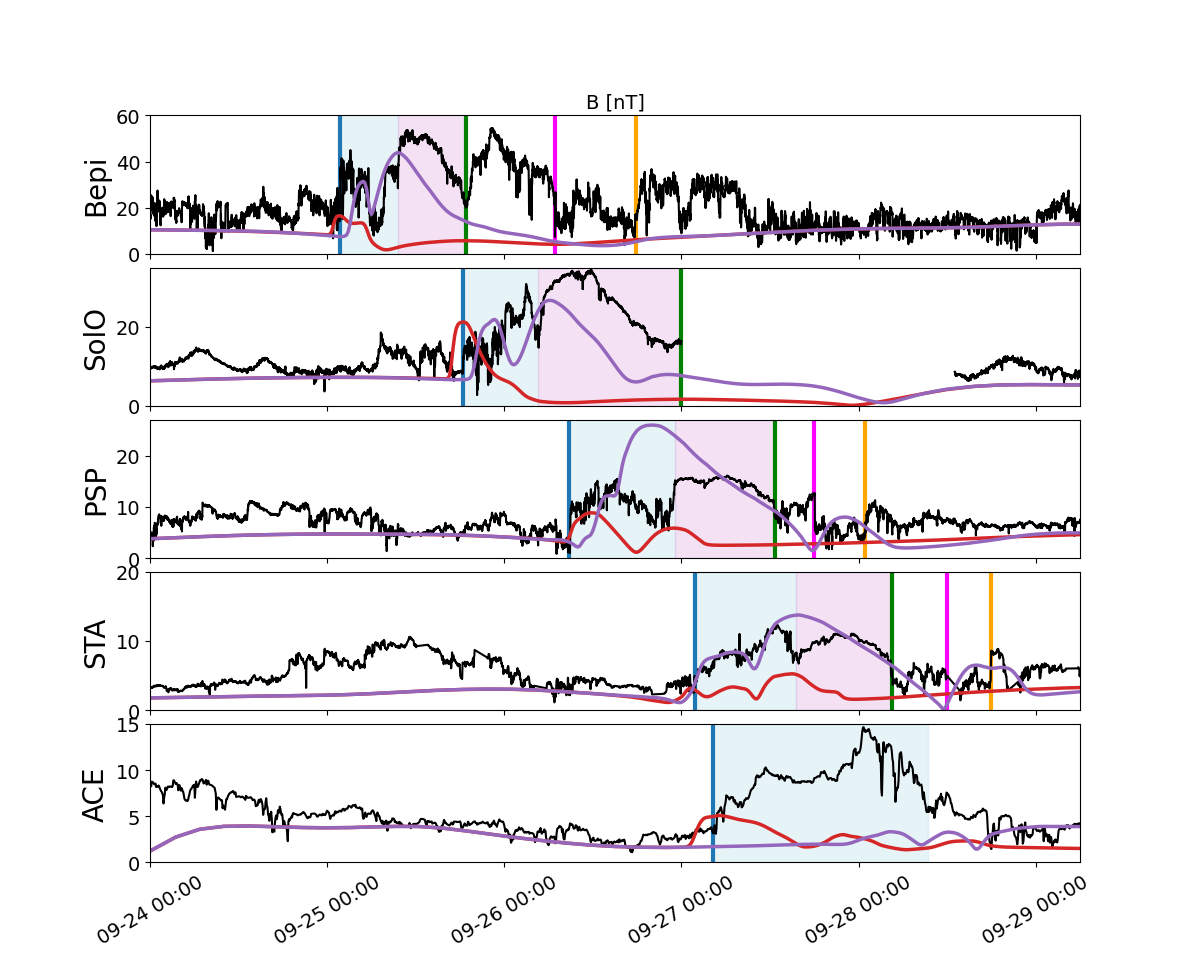}
    \hfill
    \includegraphics[width=0.49\textwidth]{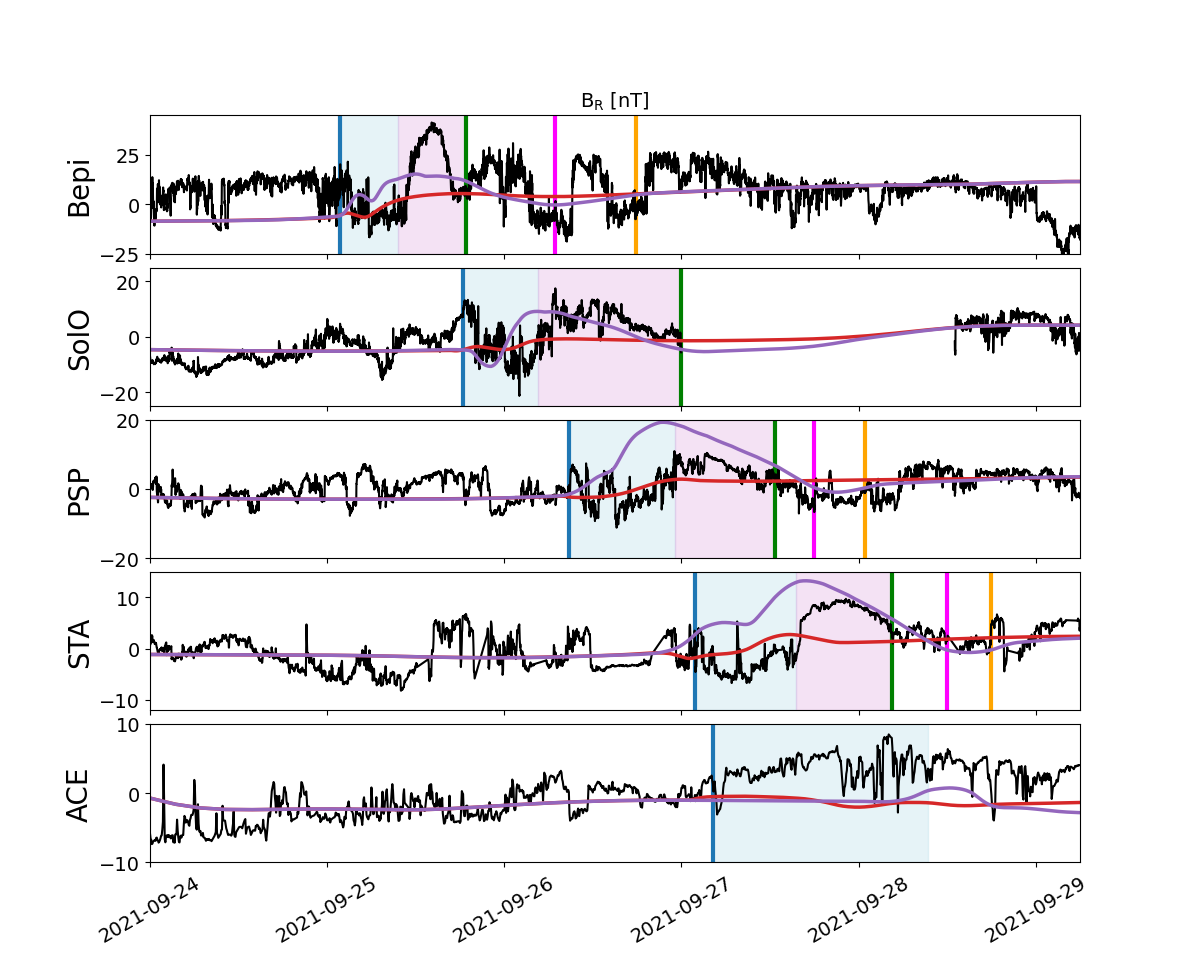}
    \hfill
    \includegraphics[width=0.49\textwidth]{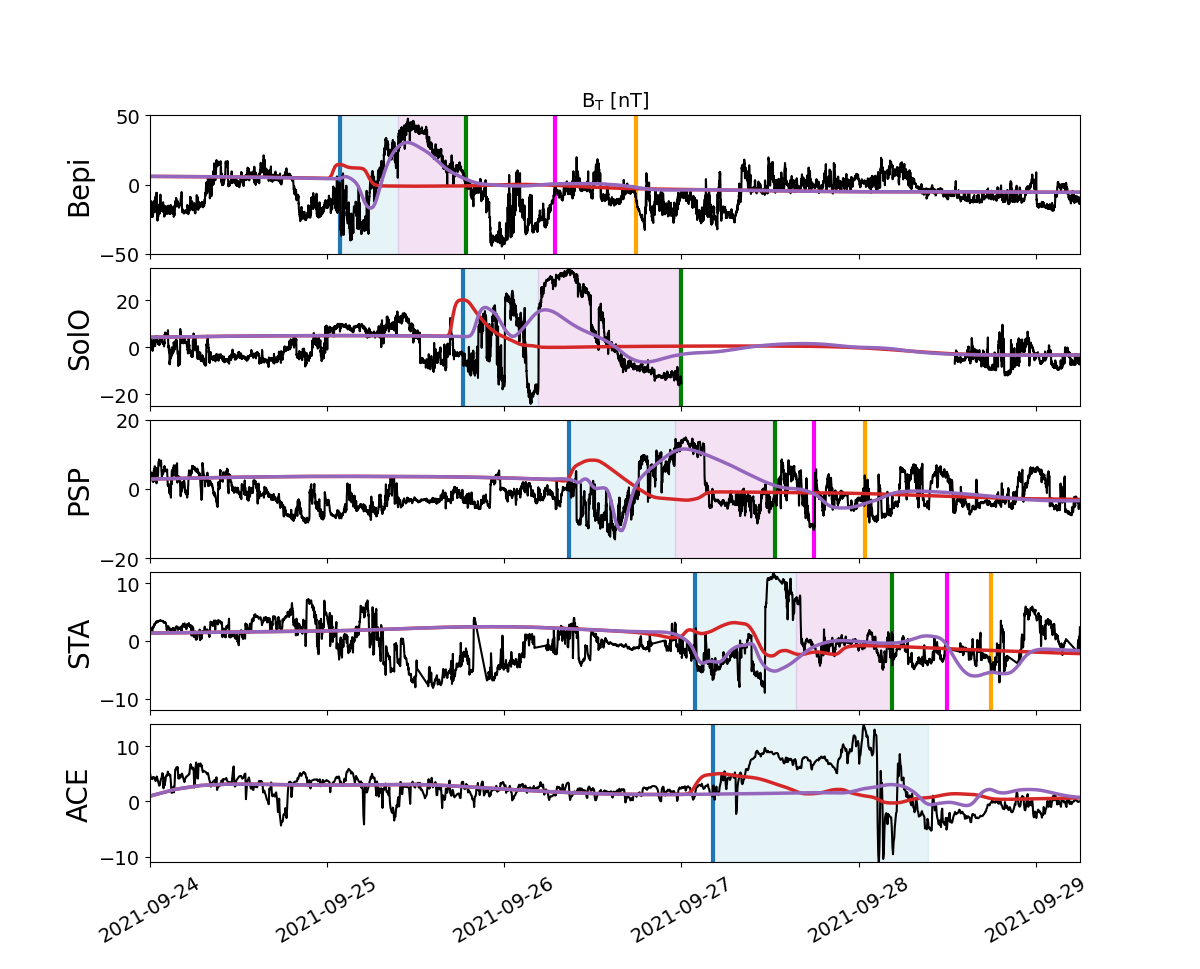}
    \hfill
    \includegraphics[width=0.49\textwidth]{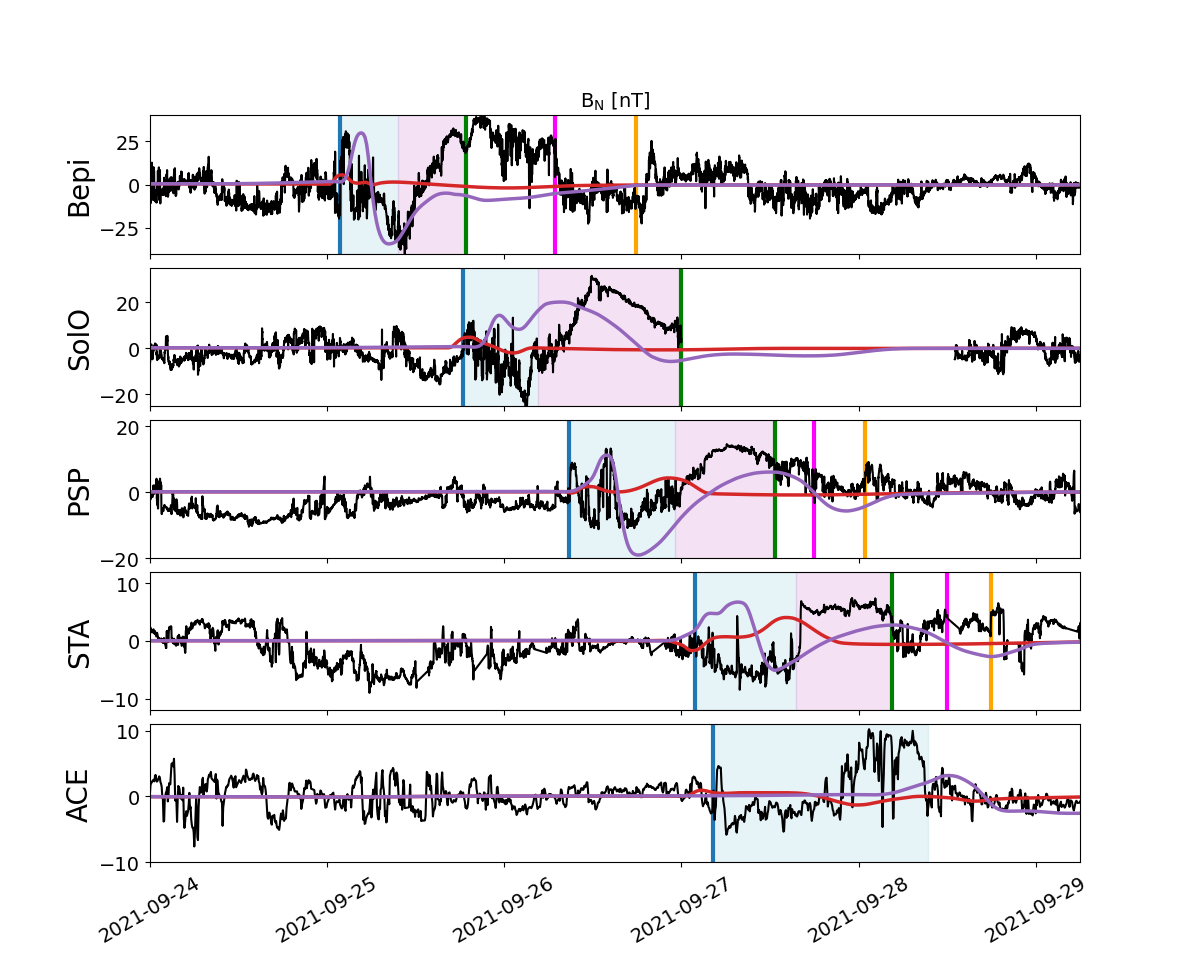}
  \caption{Magnetic field magnitude and magnetic field components in RTN coordinates at each spacecraft. The observational data are plotted in black, and the two CME models propagated in the steady solar wind, namely a cone and a spheromak CME model, are represented by the red and purple curves, respectively. The blue and pink shaded areas correspond to the sheath and magnetic ejecta regions, as defined by in situ observations. The vertical blue line shows the arrival of the shock, the vertical green line corresponds to the end of the magnetic cloud of the CME, the magenta line corresponds to the end of the second magnetic cloud, and the vertical orange line shows the end of a part of the solar wind following the CME.
  } \label{fig:all_sats_B}
\end{figure*}

\subsection{CME parameters}

The CME parameters were based on the findings of \cite{Palmerio2025} and further optimised for each CME model to {better} match the observations. Figure~\ref{fig:LASCO_C2_C3} shows the base difference of LASCO C2 and C3 coronagraph images. The photospheric HMI magnetogram is shown on the solar disk. The two green arrows show the CME extent in the plane of the sky. We considered the two parts, separated by a small darker patch in the image projection, as a single CME structure, while \cite{Palmerio2025} did not consider the smaller portion propagated southward (see Fig. 3 in \cite{Palmerio2025}). The structure propagates coherently from the solar atmosphere throughout the whole field of view of C2+C3. 

Two CME models were used in the simulations. The cone CME model \citep{Scolini2018} introduces a simple non-magnetised hydrodynamic plasma cloud in the domain. The linear force-free spheromak \citep{Verbeke2019} represents a more advanced flux rope structure with an internal magnetic field configuration. In total, four heliosphere simulations were performed because the two CME models were propagated in the steady and dynamic solar wind backgrounds.

Table~\ref{table:CME_parameters} summarises the parameters we used in the Icarus simulations for injecting the cone and spheromak CME models. The injection time at 0.1~au ({t$_{\rm CME}$}), angular location ($\theta_{\rm CME}$, $\phi_{\rm CME}$) and size (r$_{\rm CME}$) were identical for both models. The difference in the injection speed (v$_{\rm CME}$) arises from the overexpansion problem characteristic of the spheromak model \citep{Scolini2019}. The temperature of the CME is described by the T$_{\rm CME}$ parameter. Next, $\tau_{\rm CME}$, H$_{\rm CME}$, and F$_{\rm CME}$ describe the tilt, the magnetic helicity sign, and the magnetic flux of the spheromak model, respectively ($\tau_{\rm CME}$, H$_{\rm CME}$, and F$_{\rm CME}$ were not used in the cone model). 

\section{Simulation results} \label{sec:Results}

\begin{figure*}[hpt!]
\centering
    \includegraphics[width=\textwidth]{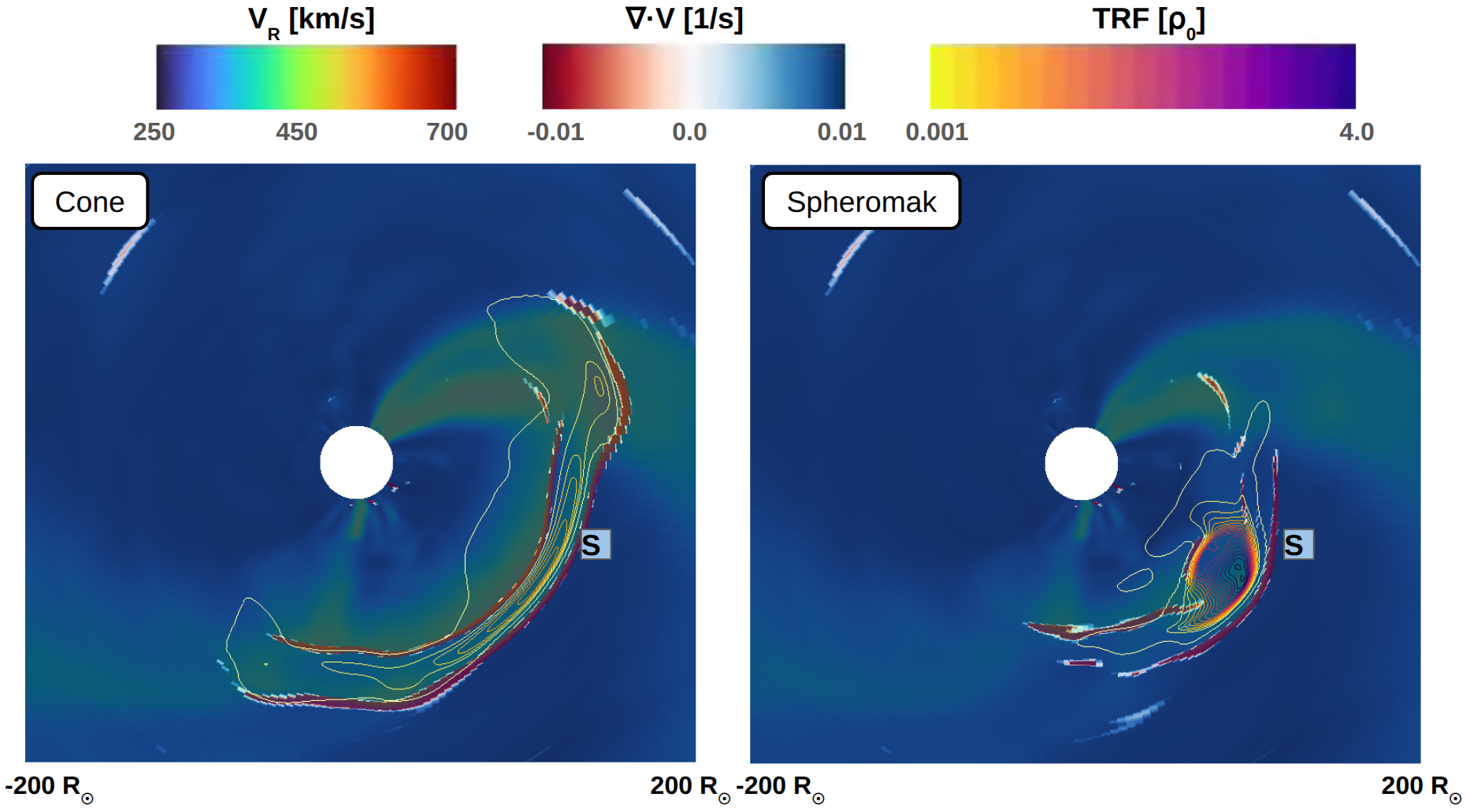}
  \caption{Equatorial planes in the 3D dynamic heliosphere simulations with the modelled cone (left) and spheromak (right) CMEs. The radial velocity and velocity divergence are plotted in the background. The divergence of velocity is saturated to show compression regions in dark red. Outside these compression regions, the colour function of the velocity divergence is faded so that the radial velocity is shown. The contour of the CME tracing function, TRF, is overplotted. 
  The snapshot corresponds to the observed arrival time of the shock at Solar Orbiter. The location of Solar Orbiter is indicated by `S'.}
  \label{fig:cone_spheromak_divv_v_trp_SolO}
\end{figure*}

\subsection{Comparison with in situ magnetic field measurements}
The signatures of the CME structure were observed at various distances from the Sun by five spacecraft. Figure~\ref{fig:all_sats_B} follows the CME structure as it travelled outwards through the heliosphere. %Each panel displays the same variable time series observed by the five spacecraft, shown in black: the total magnetic field strength and the magnetic field components in the radial-tangential-normal (RTN) coordinate system. All variables are given in nT. A blue vertical line indicates the shock arrival at each spacecraft; the green vertical line corresponds to the end of the magnetic cloud of the CME studied in this paper, the magenta line corresponds to the end of the second magnetic cloud {(introduced later in the text)}, and the orange vertical line shows the end of a part of the solar wind following the CME. The shaded blue and pink areas represent the CME sheath and magnetic ejecta regions, respectively. 

The two simulation results with the cone and spheromak models are plotted in red and purple, respectively. The two CMEs were propagated in the steady solar wind to facilitate comparison. Overall, the magnetic field modelled by the spheromak mimics the observed solar wind fairly well, while the magnetic field in the case of the cone CME model is poorly modelled. This is expected because the cone CME model lacks an intrinsic magnetic field and only perturbs the solar wind's magnetic field as it propagates. 

The modelled magnetic field of the solar wind immediately before the arrival of the CME at the spacecraft is not accurate and is mostly underestimated in the Icarus simulation results. The total magnetic field of the CME is modelled well with the spheromak model, except for PSP, where the total magnetic field strength is overestimated and for ACE, which missed the magnetic cloud. Elsewhere, the double-peak profile that corresponds to the increase in the magnetic field strength in the CME sheath and the magnetic cloud, is reconstructed accurately. 

At the location of BepiColombo, we can infer from the observations that two magnetic clouds interact with each other. The magnetic field of the second magnetic cloud fluctuates, but still displays some degree of rotation.  We were unable to confirm the second magnetic cloud at Solar Orbiter because of a data gap, but its traces are present at PSP (Figure~\ref{fig:PSP_B_plasma}) and STA (Figure~\ref{fig:all_sats_B}), with some difference in the magnetic field strength profile that probably arise from the different regions that were crossed and from the expected evolution with heliocentric distance.  
At PSP and STA, the magnetic field of the second magnetic cloud still rotated. Its velocity is comparable to the velocity at the end of the first magnetic cloud, and there is no evidence of compression. Then, the second magnetic cloud is mostly stacked behind the first cloud with negligible forces involved. It is therefore expected, at least at these heliocentric distances, to have a negligible effect on the propagation of the first cloud. Behind the second magnetic cloud, a slightly faster solar wind is present, with an enhanced velocity at STA compared to PSP. There is indeed clear evidence of compression at STA with localised peaks in the magnetic field strength and density (around the orange line). This stream is likely located at the border of the fast stream observed at ACE.
Finally, we conclude that the second magnetic cloud is stacked behind the first cloud, so that it does not significantly affect its evolution, while it partially protects the first cloud from the fast solar wind stream on its western side.

The values of the radial magnetic field component are underestimated at BepiColombo in the spheromak simulations compared to observed values, but are well modelled at Solar Orbiter. At PSP and STA, they are slightly overestimated compared to the observations. The overall profiles match the observations, however. 

The tangential component of the magnetic field is well modelled at BepiColombo. The dip observed between the sheath and the magnetic cloud is poorly modelled at Solar Orbiter. The sheath is accurately replicated at PSP by the spheromak simulation, and the magnetic cloud B$_T$ is also similar to observations. Instead of the abrupt reversal of the polarity, but B$_T$ gradually decreases to negative values. The second part of the sheath is modelled poorly at STA. Here, flank encounter of the CME with the spacecraft is visible. The sheath is observed to be longer than the magnetic cloud at STA. 
The computed profiles do not follow the observed ones in the sheath, but the solar wind values are recovered after the sheath.

\begin{figure*}[hpt!]
\centering
    \includegraphics[width=0.49\textwidth]{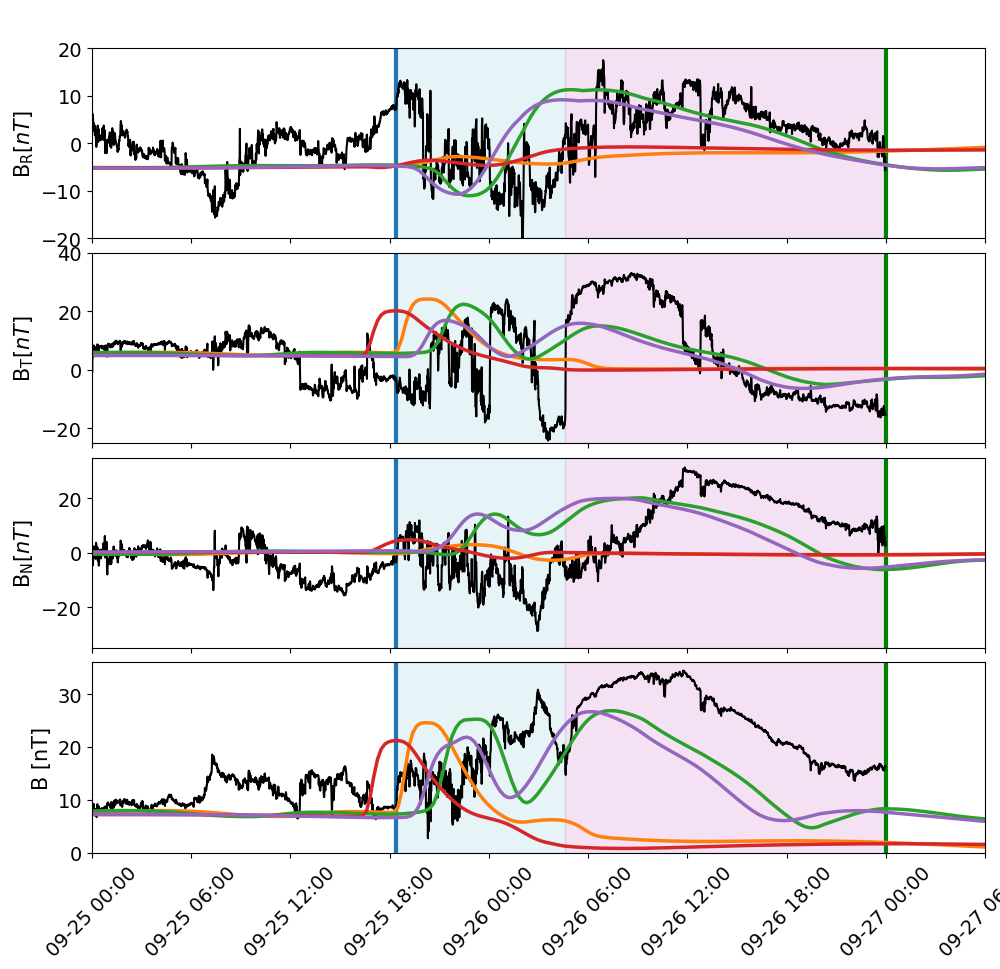}
\includegraphics[width=0.49\textwidth]{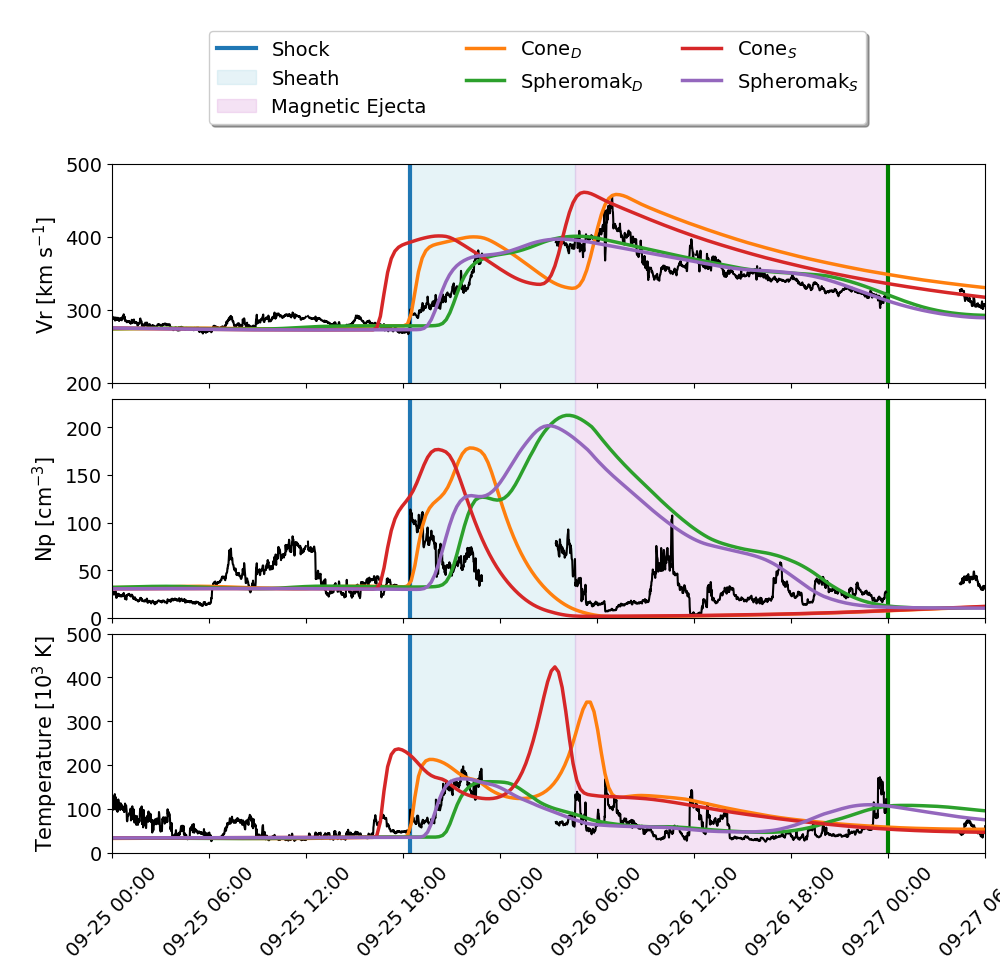}
  \caption{Time series of the magnetic (left) and plasma (right) variables at Solar Orbiter. The shock is indicated with the vertical blue line. The green line denotes the passing of the magnetic ejecta. The blue and pink shades indicate the sheath and the regions of magnetic ejecta. The dynamic solar wind simulations, using the cone and spheromak CME models, are depicted in orange and green, respectively. The steady solar wind simulations with cone and spheromak CMEs are shown in red and purple. } \label{fig:SolO_B_plasma}
\end{figure*}

The B$_N$ component {in the sheath} is modelled well by a spheromak model at BepiColombo and PSP. The magnetic cloud is underestimated at BepiColombo and recovered well at PSP. The sheath is not modelled well at Solar Orbiter, which also affects the magnetic cloud. The magnetic cloud region is modelled better at STA. The mild disturbances in the time series at ACE arrive later in the simulation than in the observations, and the signature is strongly underestimated. 

At ACE, two successive high-speed streams in the solar wind are observed immediately before and after the arrival of the CME sheath at L1. The WSA-generated boundary conditions do not reconstruct this region well, however, and the high-speed streams are not modelled correctly (see Section \ref{appendix_ACE} and Figure~\ref{fig:L1_B_plasma}). This might affect the sheath of the propagating CME and its modelling at L1. At L1, the computed disturbance arrives much later, at about the end of the observed sheath. The cone model is generally unsuitable for investigating the evolution of the magnetic structure of a CME throughout the heliosphere. This was expected because the slight variations in the magnetic field components are solely due to the solar wind signatures that are affected by the propagating plasma cloud.

\subsection{Detailed comparison at Solar Orbiter}
The two CME models were injected from the inner heliosphere boundary in the steady and dynamic solar wind regimes. In this subsection, we select Solar Orbiter observations to compare observations and simulations in more detail. Figure~\ref{fig:cone_spheromak_divv_v_trp_SolO} shows the cone (left) and spheromak (right) models propagating in the dynamic solar wind in the equatorial plane from the Icarus simulation. The snapshot was taken at the observed shock arrival time at Solar Orbiter from Table~\ref{table:cme_arrival_times}. The radial velocity is plotted in the background, overlaid with the $\nabla \cdot \Vec{V}$ values, which are saturated so that dark red regions indicate compression regions, thus indicating shocks within the domain. The contours of the CME tracing function (TRF) are plotted for the cone and spheromak models. The TRF is an in-built functionality in \texttt{MPI-AMRVAC}, showing only the CME density values, with zero everywhere else in the domain. The corresponding colour maps are provided at the top. The snapshot is zoomed in to show $\pm200\;$R$_\odot$ from the Sun. The location of Solar Orbiter is also marked in the panels. 

The numerical results show that the cone CME {has} expanded in the longitudinal direction and is somewhat compressed in the radial direction. The spheromak model shows the opposite; it {has} expanded less in the longitudinal direction and is not compressed as much in the radial direction. The cone and spheromak models evolve differently, even thought they were introduced with the most similar parameters and interact with the same solar wind. This difference in evolution mainly arises from the magnetic forces in the spheromak model. First, the magnetic tension limits the longitudinal extension. 
Second, the internal magnetic pressure induces an expansion in the radial direction \citep{DemoulinDasso2009}. These effects are stronger because the spheromak Alfv\'en crossing time is shorter than the CME propagation (or expansion) time \citep{Sangalli2025}.

\begin{table}[ht!]   
\centering  
\caption{Arrival times for {the observed CME shock and} for the cone and spheromak CME models. in the steady (S) and dynamic (D) solar wind regimes at Solar Orbiter.}   
\begin{tabular}{lc}         
\hline 
 & Arrival time (UT) \\ 
\hline  \hline        
    {Observed} & {2021-09-25T18:25:00} \\ 
    Cone$_D$ & 2021-09-25T18:16:00  \\ 
    Cone$_S$ & 2021-09-25T16:19:00 \\ 
    Spheromak$_D$ & 2021-09-25T20:42:00 \\
    Spheromak$_S$ & 2021-09-25T19:25:00\\ 
\hline                           
\end{tabular}
\tablefoot{$\theta,\phi$ denote longitude and latitude, respectively. }
\label{table:icarus_arrival_solo}
\end{table}

\begin{figure*}[hpt!]
\centering
    \includegraphics[width=\textwidth]{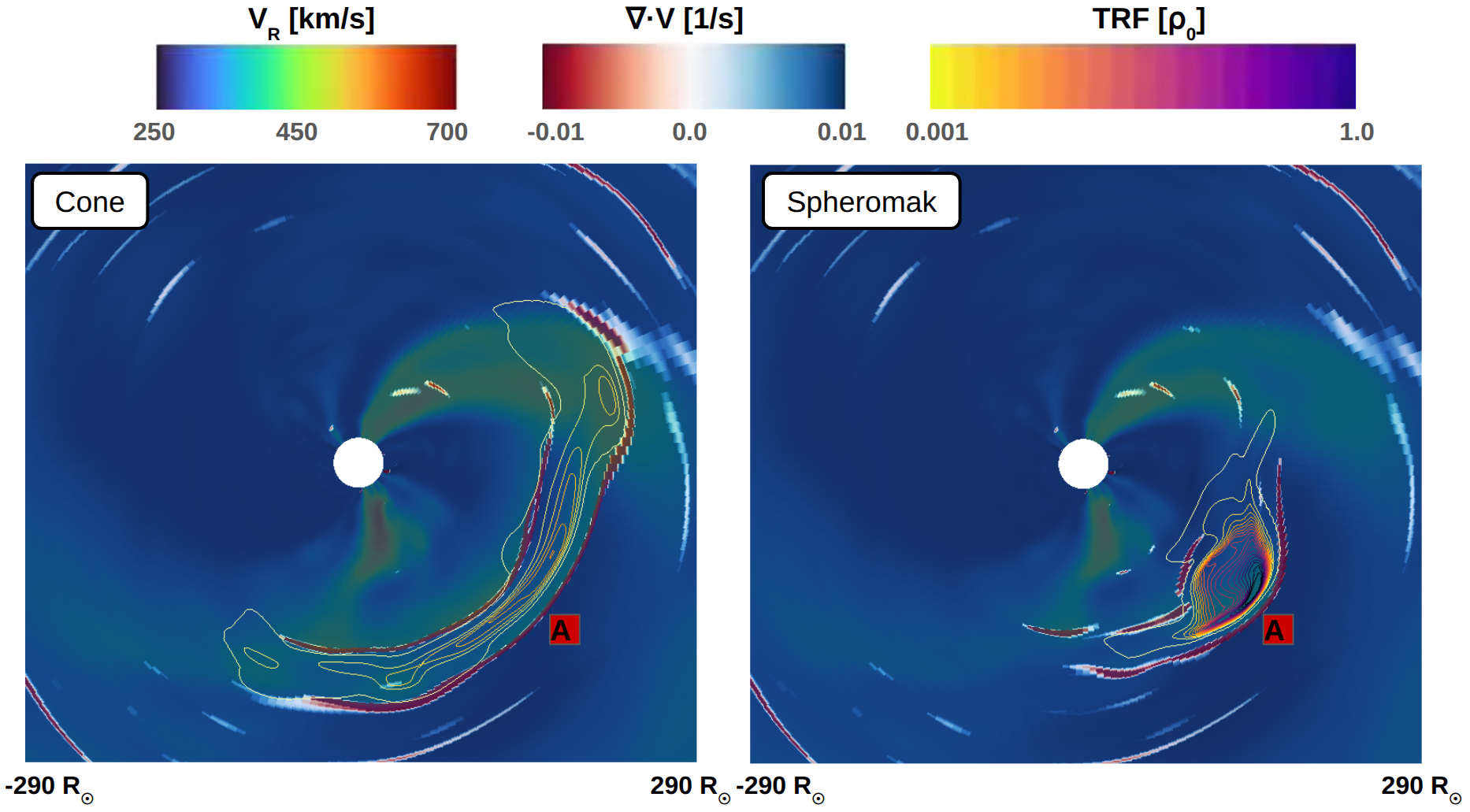}
  \caption{Equatorial planes in the 3D dynamic heliosphere simulations with a cone and spheromak CMEs. The figure is presented in the same format as Fig.~\ref{fig:cone_spheromak_divv_v_trp_SolO}. The snapshot corresponds to the observed arrival of the shock at STA. The STA location is indicated in the figure by `A'.} \label{fig:cone_spheromak_divv_v_trp_STA}
\end{figure*}
\begin{figure*}[hpt!]
\centering
    \includegraphics[width=0.49\textwidth]{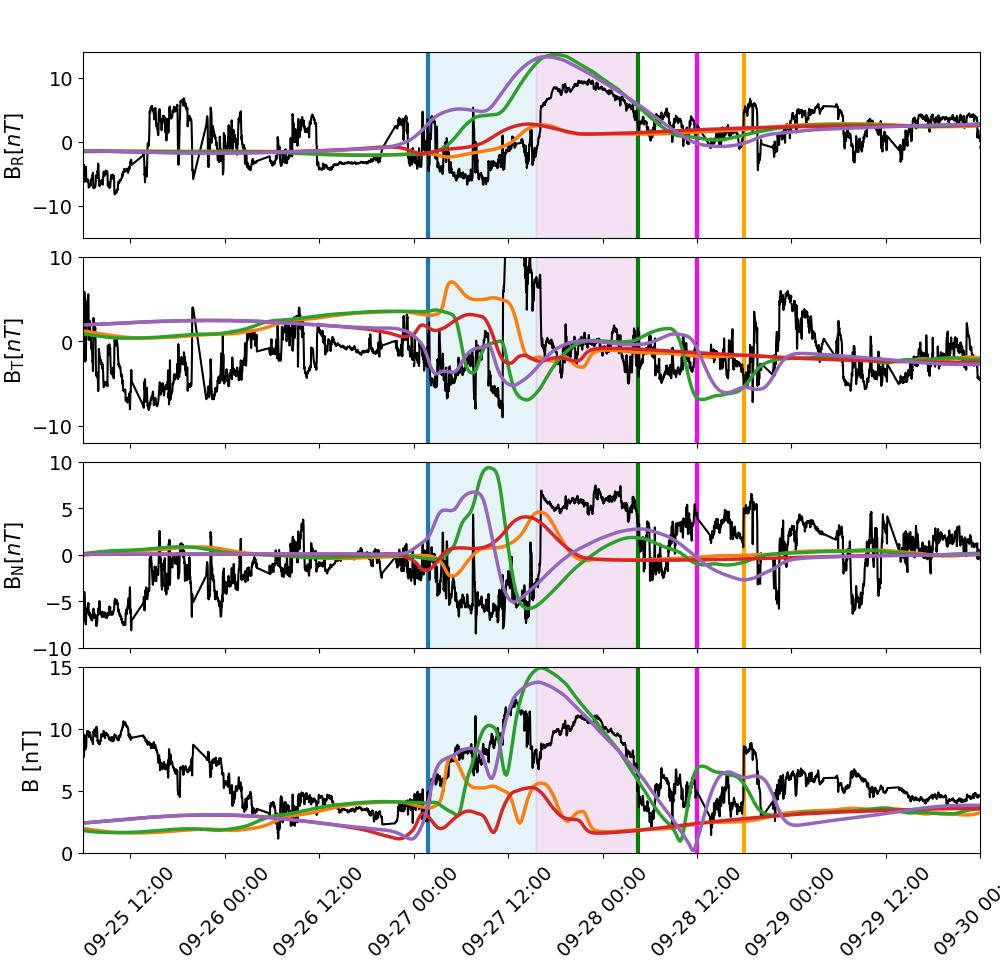}
    \includegraphics[width=0.49\textwidth]{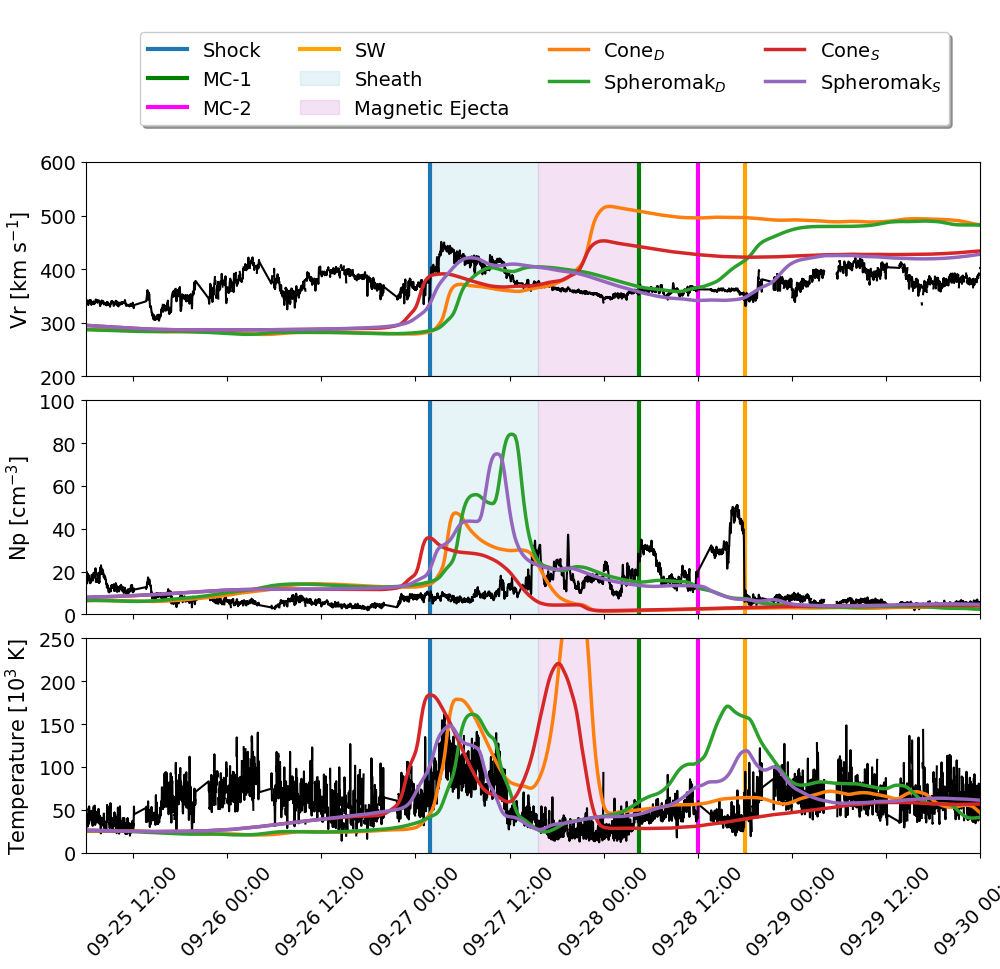}
  \caption{Time series of the magnetic (left) and plasma (right) variables at STA. The figure is presented in the same format as Fig.~\ref{fig:SolO_B_plasma}. The vertical blue line shows the arrival of the shock, the vertical green line corresponds to the end of the magnetic cloud of the CME, the magenta line corresponds to the end of the second magnetic cloud, and the vertical orange line shows the end of a part of the solar wind following the CME.
  } \label{fig:STA_B_plasma}
\end{figure*}

Table~\ref{table:icarus_arrival_solo} shows the arrival times of {the observed CME shock} and the cone and spheromak CME models in the steady and dynamic solar wind regimes. The observed arrival of the shock is best reproduced by the cone CME propagated in the dynamic solar wind. The cone CME model, propagating in the steady and dynamic solar wind, arrives earlier than the spheromak model. Moreover, in both cases, the CME model propagating in the dynamic solar wind arrives later than the CME propagating in the steady solar wind. The cone CME arrives $\sim2$ hours later in the dynamic solar wind regime than in the steady solar wind, and the spheromak in the dynamic solar wind arrives $\sim1.25$ hours later than in the steady solar wind. 

Furthermore, we investigated the global propagation of simulated CMEs and examined magnetic and plasma variables. Figure~\ref{fig:SolO_B_plasma} shows the observed and modelled data at Solar Orbiter. The velocity and temperature profiles in Figure~\ref{fig:SolO_B_plasma} (right panels) are better reconstructed by the spheromak model. The second jump in velocity for the cone model is model related, and no such profile is observed in the spheromak CME model. Unfortunately, the CME sheath is only partially observed by the plasma instrument onboard Solar Orbiter. The global structure of the magnetic and plasma profiles is still reconstructed by the spheromak model in agreement with the observations. 
The CME sheath is slightly more extended and characterised by a stronger total magnetic field strength in the dynamic solar wind case. Compared to in situ observations, the plasma density is poorly modelled, independent of the CME model. Overall, when the spheromak CME model is propagated in the dynamic solar wind, the observational signatures are reconstructed best at Solar Orbiter, with only minor differences compared to the spheromak CME model in the steady solar wind. 

\subsection{Global CME evolution: detailed comparison at STA}
Next, we considered the 3D results, and the magnetic and plasma time series at STA. STA crossed a different portion of the CME (see Figure~\ref{fig:steady_vs_dynamic_wind} for the spacecraft positions).  
Figure~\ref{fig:cone_spheromak_divv_v_trp_STA} is arranged in the same way as Figure~\ref{fig:cone_spheromak_divv_v_trp_SolO}, indicating the location of STA at the observed shock arrival time at STA as given in Table~\ref{table:cme_arrival_times}. Here, the domain was extended to $\pm 290$ R$_\odot$ to show the whole CME structure. 
The two CME models are more extensive compared to Solar Orbiter, because they have reached 0.96~au in the domain. 

Figure~\ref{fig:STA_B_plasma} shows the magnetic and plasma quantities at STA. The figures are arranged similarly to Figure~\ref{fig:SolO_B_plasma}. The spheromak simulations overestimate the radial component of the magnetic field. The first part of the sheath is modelled well by spheromak simulations, but the cone model reproduces the opposite polarity in the $B_T$ component. The sheath is modelled poorly in the $B_N$ component by the two models, but the $B_N$ profile within the magnetic ejecta matches the observations better. The two models fail to capture the peak of $|B|$ in the second part of the sheath. For all the magnetic field components, the arrival time of the spheromak propagated in the steady solar wind agrees better with the observations. The sheath is also slightly more compressed in the dynamic case, and it arrives later. 

\begin{table}[ht!]   
\centering  
\caption{Arrival times {for the observed CME shock and} for the cone and spheromak CME models in the steady ($S$) and dynamic ($D$) solar wind regimes at STA.}   
\begin{tabular}{lc}         
\hline 
 & Arrival time (UT) \\ 
\hline  \hline      
    {Observed} & {2021-09-27T01:51:00} \\
    Cone$_D$ & 2021-09-27T03:05:00  \\ 
    Cone$_S$ & 2021-09-26T22:17:00 \\ 
    Spheromak$_D$ & 2021-09-27T03:15:00 \\
    Spheromak$_S$ & 2021-09-26T23:09:00\\ 
\hline                           
\end{tabular}
\label{table:icarus_arrival_sta}
\end{table}

Next, for the dynamic solar wind, the cone and spheromak CMEs arrive $\sim4$ and $\sim4.5$ hours later than in the steady solar wind, respectively (Table~\ref{table:icarus_arrival_sta}). 
At STA, the cone CME model arrives earlier than the spheromak CME model. For the cone and spheromak CME models, the arrival time in the dynamic solar wind is closer to the observed arrival time. The two models overestimate the proton density, and the radial velocity profiles match the observations well in the sheath. The arrival time of the cone CME model is more accurate, but the whole structure is reconstructed more accurately by the spheromak model. After the CME passes, the solar wind is similar to the observations in the steady regime; the values are overestimated in the dynamic solar wind case. The temperature profiles are also more accurate in the spheromak simulations than in the cone simulation for the steady and dynamic solar wind. 

Finally, the analysis of the in situ measurements and the 3D modelled data for BepiColombo, PSP, and ACE is given in Appendices~\ref{appendix_bepi}, \ref{appendix_psp}, and \ref{appendix_ACE}, respectively.

\section{Conclusions} \label{sec:Conclusions}
\label{sec:conclusions}
We investigated the effect of the dynamic solar wind on non-magnetised and magnetised CME model propagation in the heliosphere sampled at five spacecraft locations. The solar wind was simulated in the Icarus heliosphere model in the steady and dynamic regimes. In the dynamic solar wind case, the input boundary conditions were updated every four hours. To study the impact of the dynamic solar wind on CME propagation, we employed two CME models: a simple non-magnetised hydrodynamic plasma cloud, referred to as a cone model, and a magnetised linear force-free spheromak model.

The impact of the solar wind on the CME is a global phenomenon; it is therefore difficult to validate it only with localised observations from an individual spacecraft. To validate the model more extensively, we chose a CME event observed by multiple spacecraft spread radially and over different longitudes, within the inner heliosphere boundary and Earth. 

The CME of interest erupted on the solar surface at SOL2021-09-23T04:39:45. The source region was identified as NOAA AR 12871, and the polarity inversion line was approximated. The initial parameters were based on \citet{Palmerio2025} and were subsequently adjusted to account for the CME models we employed in the simulations. 

The evolution of the CME as it travelled through the heliosphere was investigated first. The spheromak CME model accurately replicated the magnetic cloud characteristics of the observed CME. Overall, the CME was modelled well at BepiColombo, Solar Orbiter, PSP (partially), and STA. At PSP, the total magnetic field was overestimated, primarily because the radial magnetic field component was overestimated. The tangential and normal magnetic field components were modelled better, however.
Finally, at ACE (L1), only the CME sheath was present in the observations; {signatures of the magnetic cloud were not found at this location}. 
The sheath modelled with the spheromak CME model was not strong or extended enough to reach L1, while for the cone CME model, the disturbance arrived at L1 on time, but it is too weakly compressed. 

The presented snapshots from the 3D simulations {(Figures~\ref{fig:cone_spheromak_divv_v_trp_SolO}, \ref{fig:cone_spheromak_divv_v_trp_STA}, \ref{fig:cone_spheromak_divv_v_trp_Bepi}, \ref{fig:cone_spheromak_divv_v_trp_PSP}, \ref{fig:cone_spheromak_divv_v_trp_L1})} illustrate how the two CME models evolved differently as they propagated through the heliosphere. The cone CME model tended to maintain a constant longitudinal extension as the plasma velocity remained predominantly radial and {expanded radially with a dependence slightly more than proportional with the distance to the Sun}.

In contrast, the spheromak model extends much less in longitude and expands {expanded radially with a dependence slightly less than proportional with the distance to the Sun}. This agrees better with the observed expansion of the magnetic clouds. This different evolution arises from the magnetic tension and magnetic pressure in the modelled spheromak CME. 

Another simulation result is that the two CME models arrived later at Solar Orbiter and STA when they were propagated in the dynamic, rather than steady solar wind, as shown in Tables~\ref{table:icarus_arrival_solo} and~\ref{table:icarus_arrival_sta}. This is also true for the BepiColombo and PSP given in Tables~\ref{table:icarus_arrival_bepi} and~\ref{table:icarus_arrival_psp}, respectively. 
\cite{Baratashvili2024B} reported that there was a strong deceleration for another CME observed between the orbits of Mercury and Earth, but this deceleration could not be reconstructed in the simulations using the steady solar wind solution. Using dynamic solar wind rather than steady solar wind  might enhance the results. We showed that the deceleration of the CME was greater in the dynamic solar wind  than in the steady wind.

The difference in the arrival times between the steady and dynamic solar wind cases was greater when sampled at larger in situ distances. For example, the difference at BepiColombo was $\sim2$~hours for the cone and spheromak models, while at STA, it increased to $\sim4$ and $\sim4.5$ hours for the cone and the spheromak models, respectively. We need to take into account that we cannot directly compare the differences at each spacecraft, however, because there was a significant angular separation between the spacecraft in latitude and longitude. 
At L1, we conclude that only the cone model matches the observed arrival time, regardless of the solar wind regime. 

Next, the sheath region was better modelled by the spheromak model than the cone CME model at all spacecraft, except ACE. The double profile in the total magnetic field, characteristic of the sheath and the magnetic cloud, was recovered well. Finally, the modelled sheath was less compressed in most cases compared to the observations, and it was weaker.

To conclude, the signatures observed by five spacecraft were reconstructed by the injected CME models. In most magnetic and plasma parameters, the spheromak CME model mimics the real observed data better. The longitudinal extension of the cone CME model allowed the CME to reach L1, but this was different in the spheromak CME model. In this study, the CME propagated in the slow-speed stream. Despite the homogeneous background region, the CME notably decelerated within the dynamic solar wind, with slightly affected timeline profiles. It is crucial to incorporate the dynamic solar wind background to enhance our overall forecasting capabilities and understand the sources of error in realistic solar wind scenarios. As our next step, we aim to model the CME-high speed stream interaction in the dynamic regime to demonstrate the full advantage of realistic, time-dependent solar wind modelling in the heliosphere.

\begin{acknowledgements}
Funded by the European Union. Views and opinions expressed are, however, those of the author(s) only and do not necessarily reflect those of the European Union or ERCEA. Neither the European Union nor the granting authority can be held responsible for them. This project (Open SESAME) has received funding under the Horizon Europe programme (ERC-AdG agreement No 101141362). E.~D. and E.~W. are funded by the European Union (ERC, HELIO4CAST, 101042188).

These results were also obtained in the framework of the projects
C16/24/010 C1 project Internal Funds KU Leuven), G0B5823N and G002523N (WEAVE) (FWO-Vlaanderen), and 4000145223 (SIDC Data Exploitation (SIDEX2), ESA Prodex).

We utilised the VSC – Flemish Supercomputer Centre infrastructure for the computations, which was funded by the Hercules Foundation and the Flemish Government, Department of EWI.
\end{acknowledgements}

\bibliographystyle{aa}
\bibliography{bibliography}
\begin{appendix} 
\section{In situ measurements} \label{appendix_timeseries}

    \begin{figure*}[hpt!]
\centering
    \includegraphics[width=\textwidth]{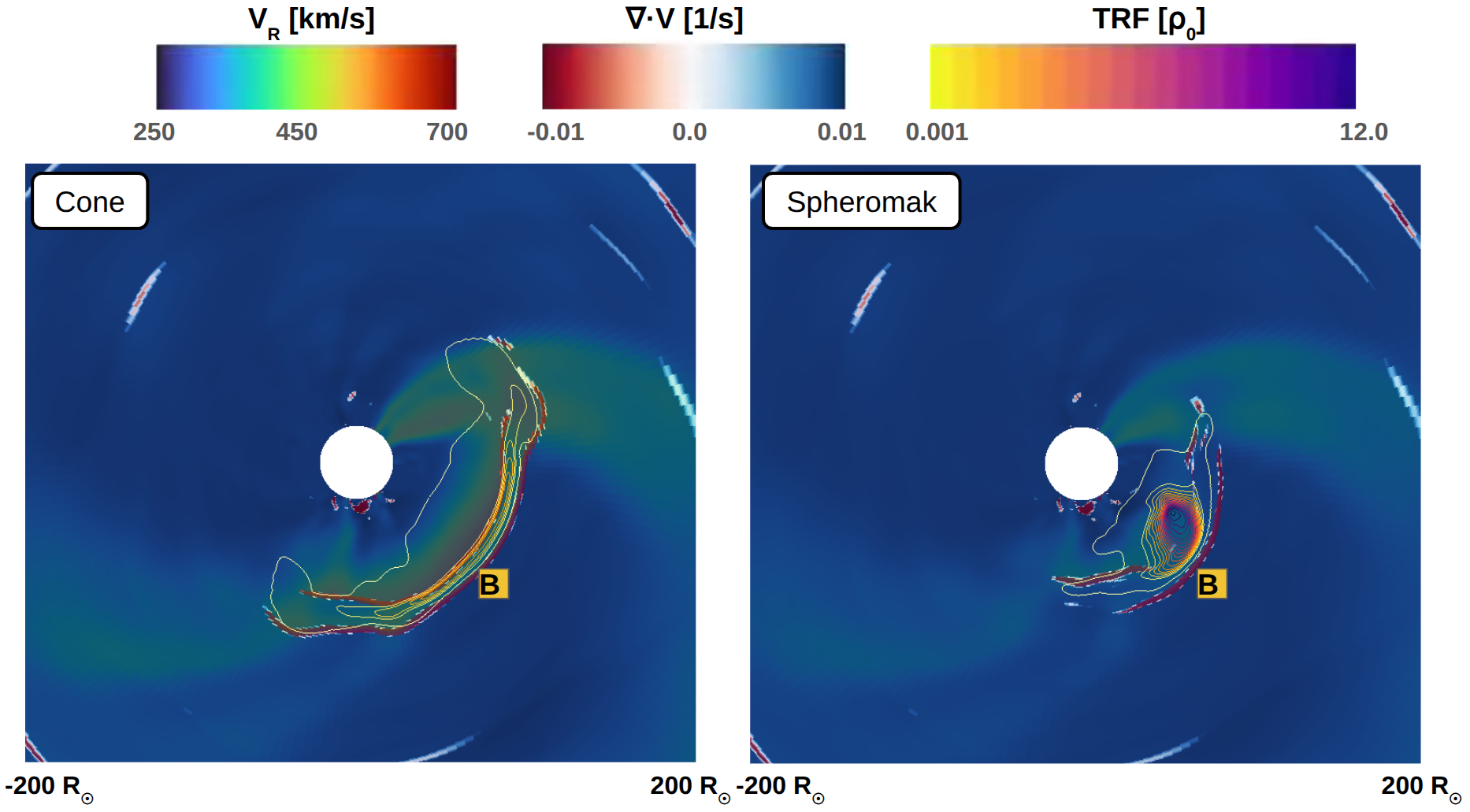}
  \caption{Equatorial planes in the 3D dynamic heliosphere simulations with the cone (left) and spheromak (right) CMEs. The radial velocity and velocity divergence are plotted in the background. The contour of the CME tracing function, TRF, is overplotted. The snapshot corresponds to the observed arrival of the shock at BepiColombo. The BepiColombo location is indicated on the figure with `B'.} \label{fig:cone_spheromak_divv_v_trp_Bepi}
\end{figure*}
\begin{figure*}[hpt!]
\centering
    \includegraphics[width=0.49\textwidth]{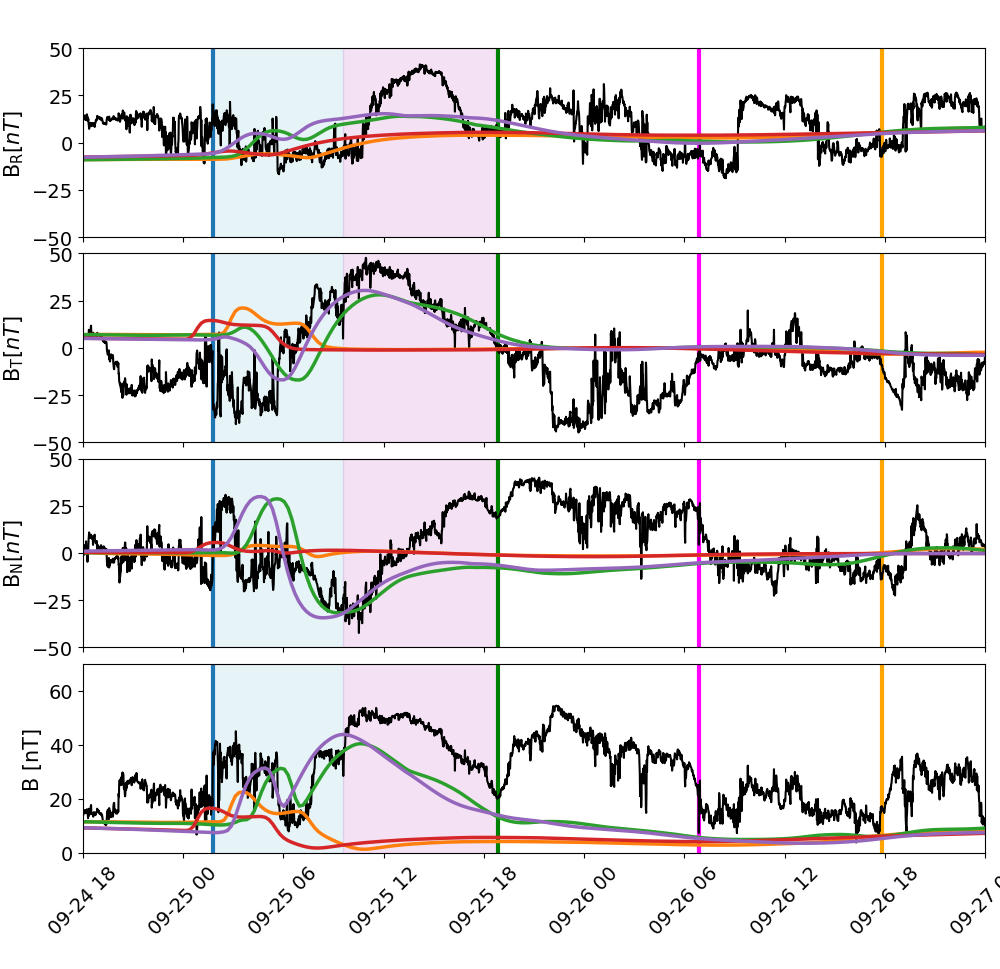}
     \includegraphics[width=0.49\textwidth]{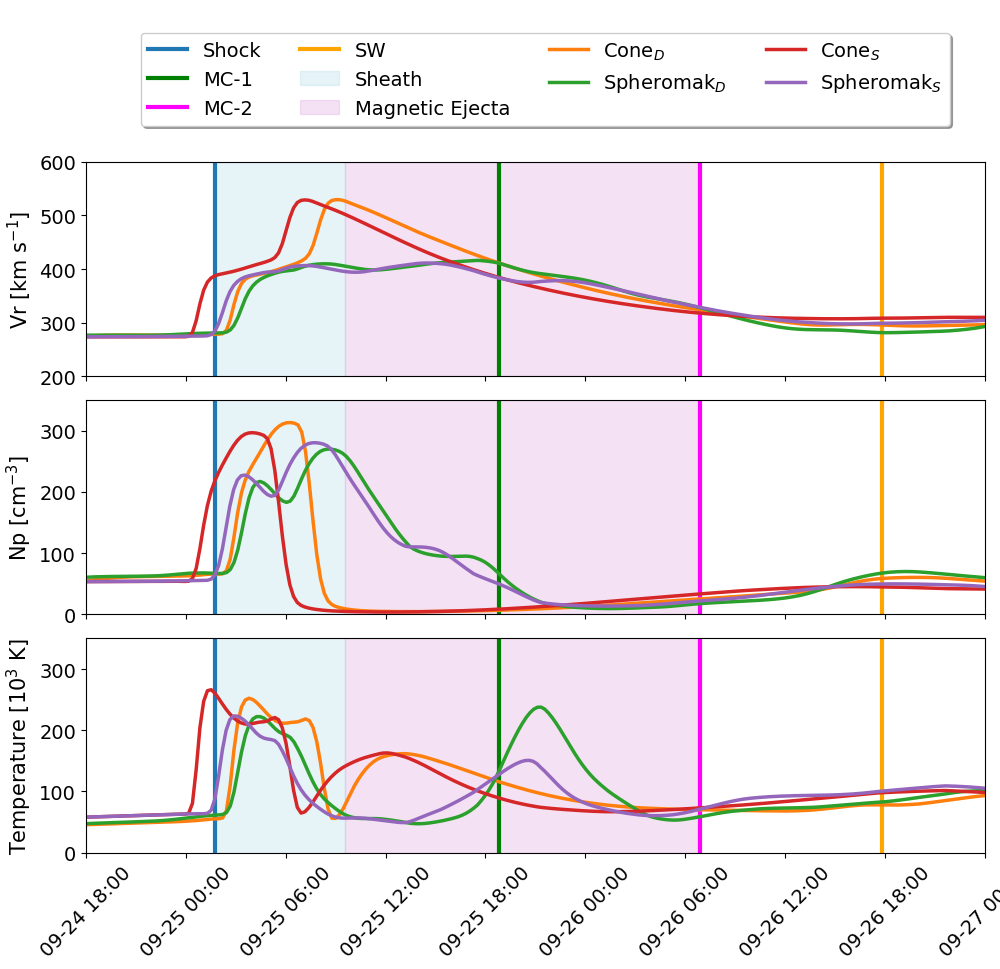}
  \caption{Timeseries of the magnetic field  {(left) and plasma quantities (right)} at BepiColombo. The blue vertical line indicates the shock, the green vertical line corresponds to the end of the magnetic cloud of the CME, the magenta line corresponds to the end of the second magnetic cloud, and the orange vertical line shows the end of a part of the solar wind following the CME. The blue and pink shaded regions indicate the sheath and the magnetic ejecta regions. The dynamic solar wind simulations, using the cone and spheromak CME models, are depicted in orange and green, respectively. Steady solar wind simulations with cone and spheromak CMEs are given in red and purple colours. The BepiColombo measurements are not plotted for plasma quantities, since these are not available from the instrument.}  \label{fig:Bepi_B_plasma}
\end{figure*}

    \begin{table}[ht!]   
\centering  
\caption{The arrival times {for the observed CME shock and} for  the cone and spheromak CME models in the steady ($S$) and dynamic ($D$) solar wind regimes at BepiColombo.} 
\begin{tabular}{lc}         
\hline 
 & Arrival time (UT) \\ 
\hline  \hline     
    {Observed} & {2021-09-25T01:46:00} \\
    Cone$_D$ & 2021-09-25T02:25:00  \\ 
    Cone$_S$ & 2021-09-25T00:22:00 \\ 
    Spheromak$_D$ & 2021-09-25T04:01:00 \\
    Spheromak$_S$ & 2021-09-25T02:40:00\\ 
\hline                           
\end{tabular}
\label{table:icarus_arrival_bepi}
\end{table}

\begin{figure*}[hpt!]
\centering
\includegraphics[width=\textwidth]{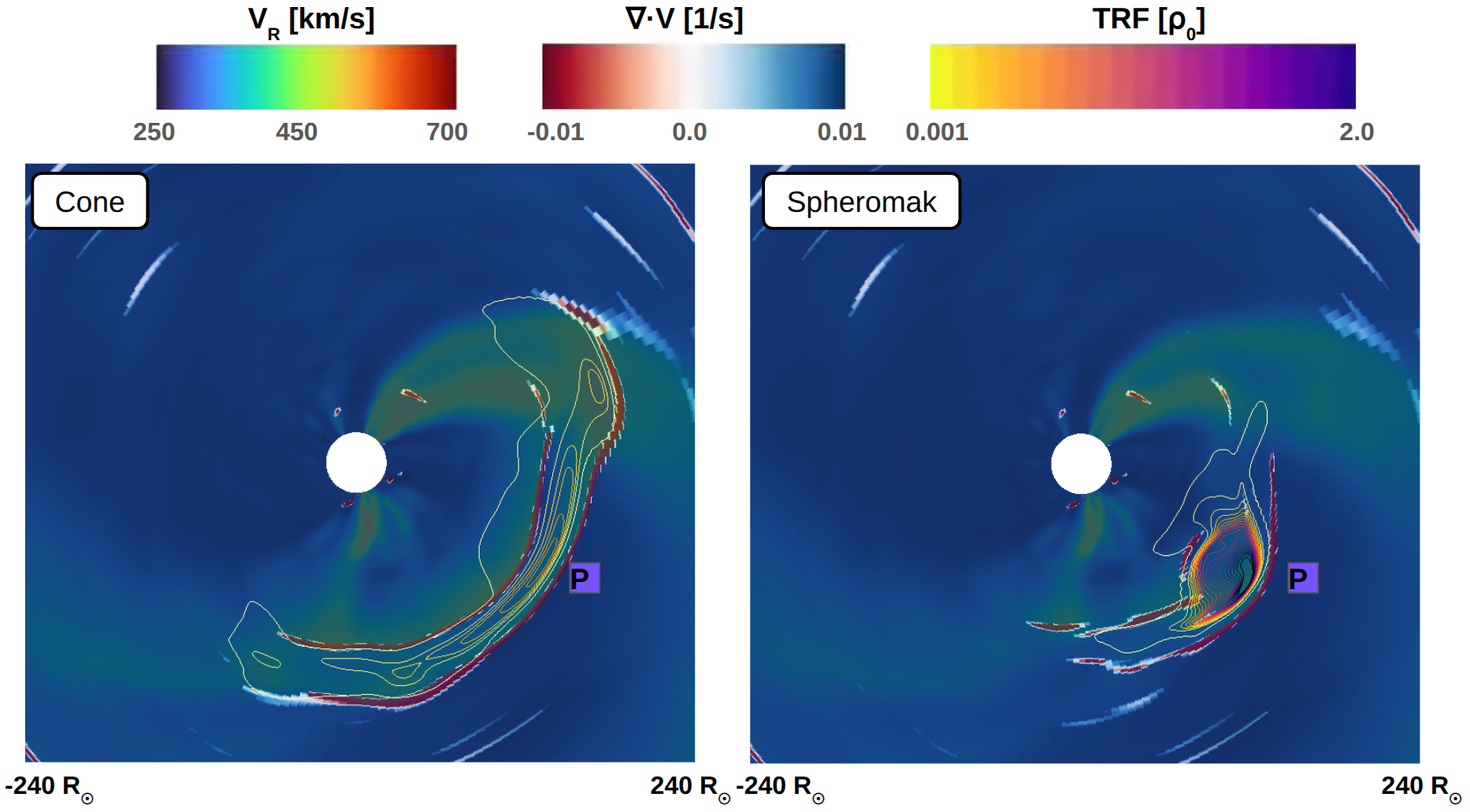}
  \caption{Equatorial planes in the 3D dynamic heliosphere simulations with cone (left) and spheromak (right) CMEs presented in the same format as Fig.~\ref{fig:cone_spheromak_divv_v_trp_Bepi}. The snapshot corresponds to the observed arrival of the shock at PSP. The PSP location is indicated on the figure with `P'.
  } \label{fig:cone_spheromak_divv_v_trp_PSP}
\end{figure*}

\begin{figure*}[hpt!]
\centering
    \includegraphics[width=0.49\textwidth]{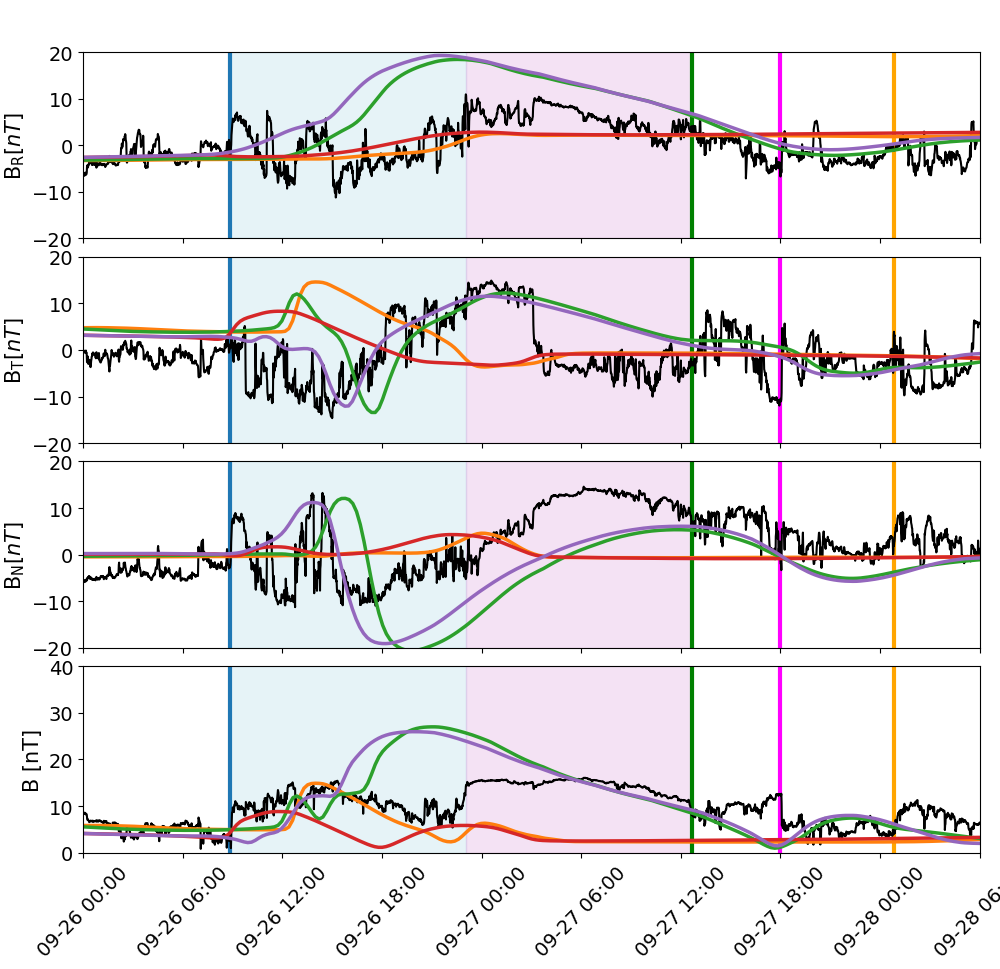}
    \includegraphics[width=0.49\textwidth]{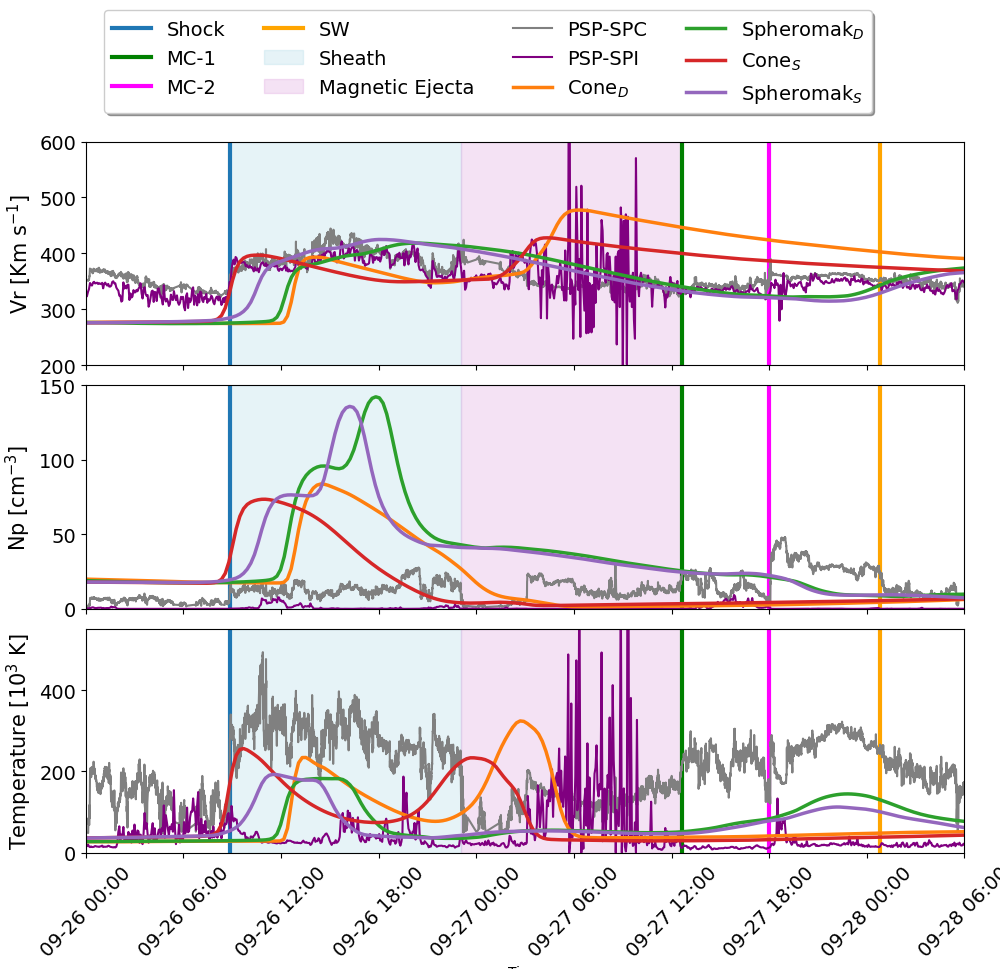}
  \caption{Time series of the magnetic (left) and plasma (right) variables at Parker Solar Probe. The figure is presented in the same format as Fig.~\ref{fig:Bepi_B_plasma}. } \label{fig:PSP_B_plasma}
\end{figure*}

\subsection{BepiColombo} \label{appendix_bepi}

Figure~\ref{fig:cone_spheromak_divv_v_trp_Bepi} shows the 2D snapshots in the equatorial plane from Icarus simulations for the cone (left) and spheromak (right) CMEs in the dynamic solar wind. The radial velocity values are plotted in the background with the corresponding colour map. The $\nabla \cdot \Vec{V}$ values are overlaid, saturated in a way that the dark red areas indicate the compression regions. The contours of the tracing function (TRF) are plotted. The location of the BepiColombo is shown in both figures. Already at 0.44~au, the cone CME is notably wider in the longitudinal direction than the spheromak CME. The significant difference in the CME shapes can already be distinguished at BepiColombo.

Figure~\ref{fig:Bepi_B_plasma} shows the in situ measurements at BepiColombo (in black) {for the magnetic field} and the Icarus simulations: a cone and a spheromak CMEs propagated in the steady and dynamic solar wind given with the subscripts $S$ and $D$, respectively. {The right panel shows the plasma quantities from Icarus simulations. The in situ plasma measurements are not plotted since these are not available.} The light blue and light pink areas indicate the sheath and magnetic ejecta regions. After the magnetic cloud of the CME studied in this paper, we can also see additional magnetic cloud signatures at Bepi. Its bounds are given with green and magenta vertical lines. The portion of the solar wind following the CME is emphasised until the orange vertical line. We can observe similar signatures later also at PSP and STA. We can see that, overall, the spheromak CME model reproduces the observed profiles better than the cone CME model. The sheath is compressed in the simulations compared to the observations. Table~\ref{table:icarus_arrival_bepi} shows the arrivals of the CMEs in the {observations and} simulations. The cone CME arrives $\sim2$ hours later in the dynamic solar wind compared to the steady solar wind, and the spheromak CME arrives $\sim1.3$ hours later in the dynamic regime compared to the steady one. {Here, the cone CME propagated in the steady dynamic solar wind, and the spheromak CME model propagated in the dynamic solar wind arrived closer to the observed arrival time compared to the other two simulations. }

\begin{table}[ht!]   
\centering  
\caption{The arrival times {for the observed CME shock and} for the cone and spheromak CME models in the steady ($S$) and dynamic ($D$) solar wind regimes at Parker Solar Probe.}   
\begin{tabular}{lc}         
\hline 
 & Arrival time (UT) \\ 
\hline  \hline       
    {Observed} & {2021-09-26T08:50:00} \\
    Cone$_D$ & 2021-09-26T12:10:00  \\ 
    Cone$_S$ & 2021-09-26T08:10:00 \\ 
    Spheromak$_D$ & 2021-09-26T11:38:00 \\
    Spheromak$_S$ & 2021-09-26T09:30:00\\ 
\hline                           
\end{tabular}
\label{table:icarus_arrival_psp}
\end{table}

\begin{figure*}[hpt!]
\centering
    \includegraphics[width=\textwidth]{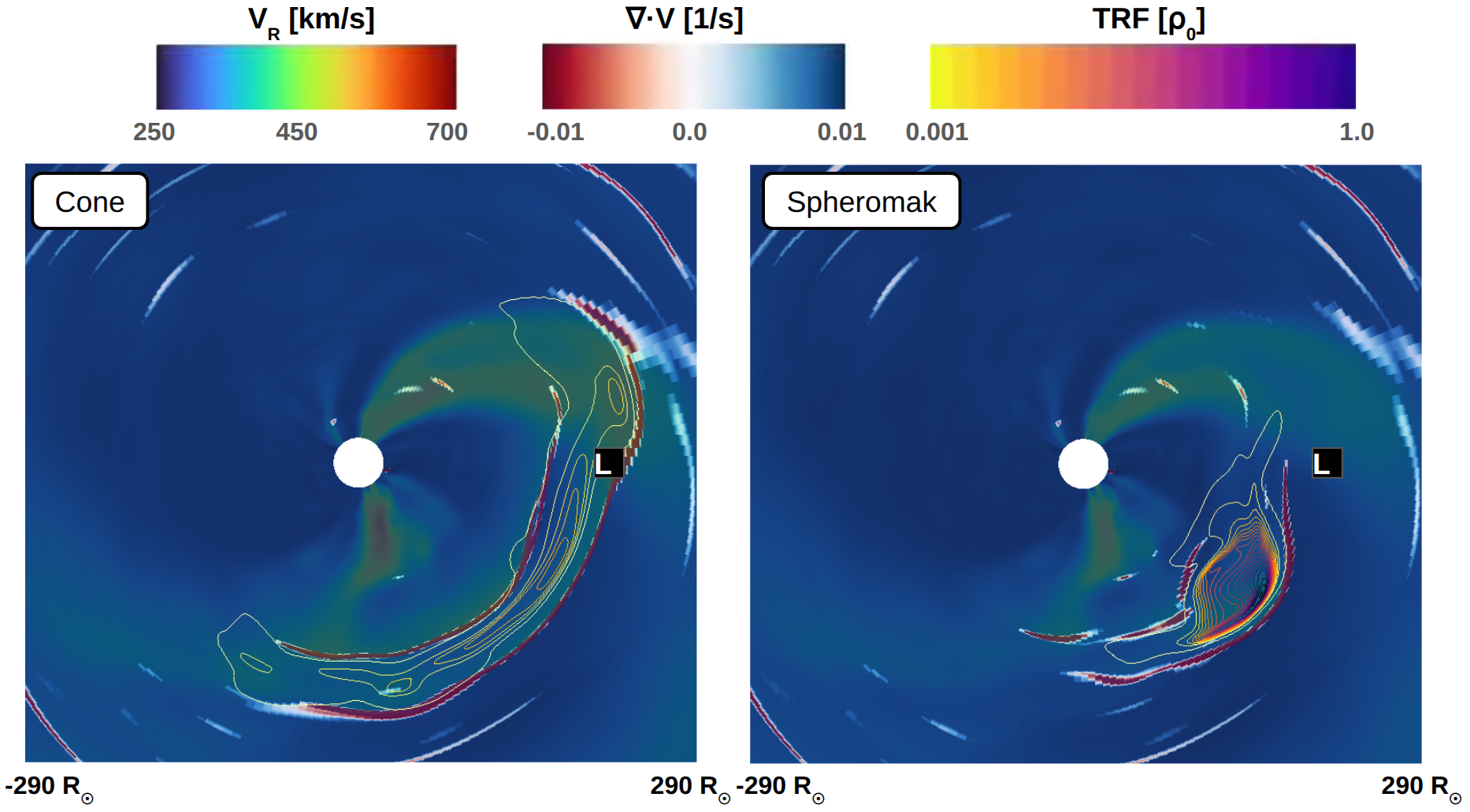}
  \caption{Equatorial planes in the 3D dynamic heliosphere simulations with cone (left) and spheromak (right) CMEs presented in the same format as Fig.~\ref{fig:cone_spheromak_divv_v_trp_Bepi}. The snapshot corresponds to the arrival of the shock at L1 (ACE). The L1 location is indicated on the figure with `L'.
  } \label{fig:cone_spheromak_divv_v_trp_L1}
\end{figure*}

\begin{figure*}[hpt!]
\centering
    \includegraphics[width=0.49\textwidth]{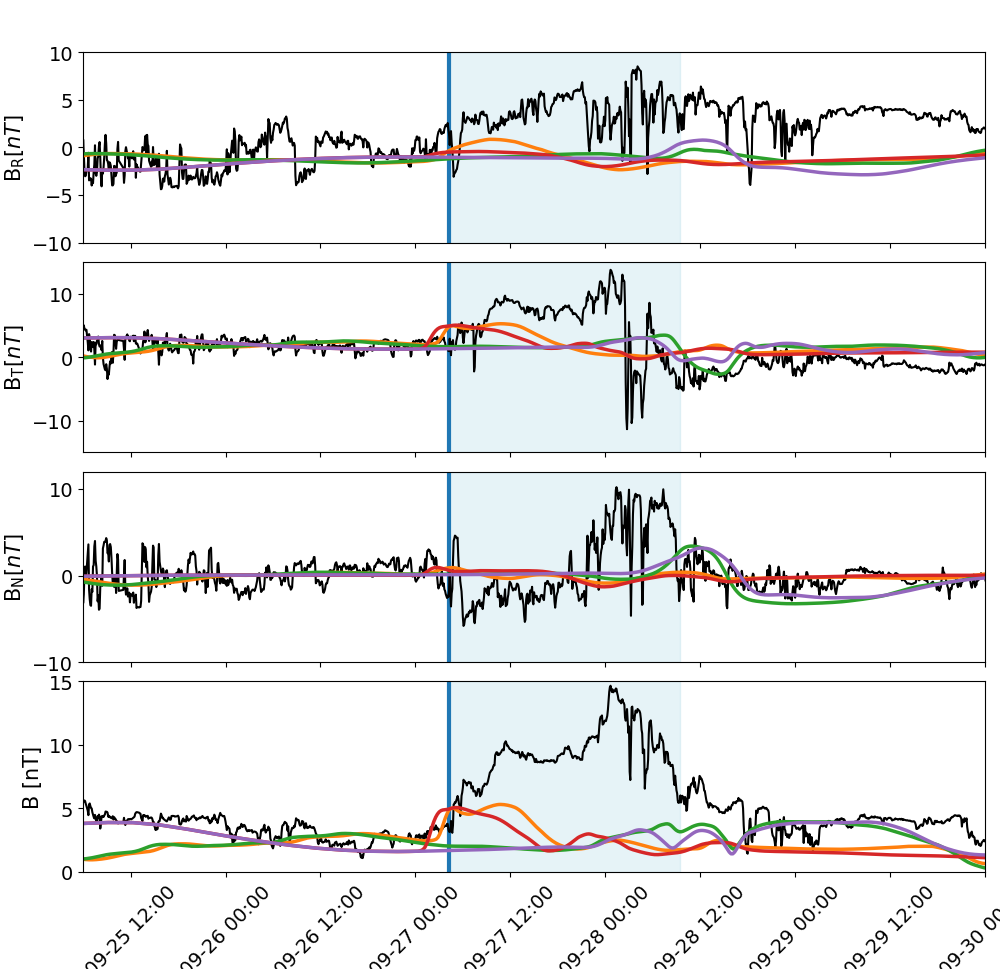}
    \includegraphics[width=0.49\textwidth]{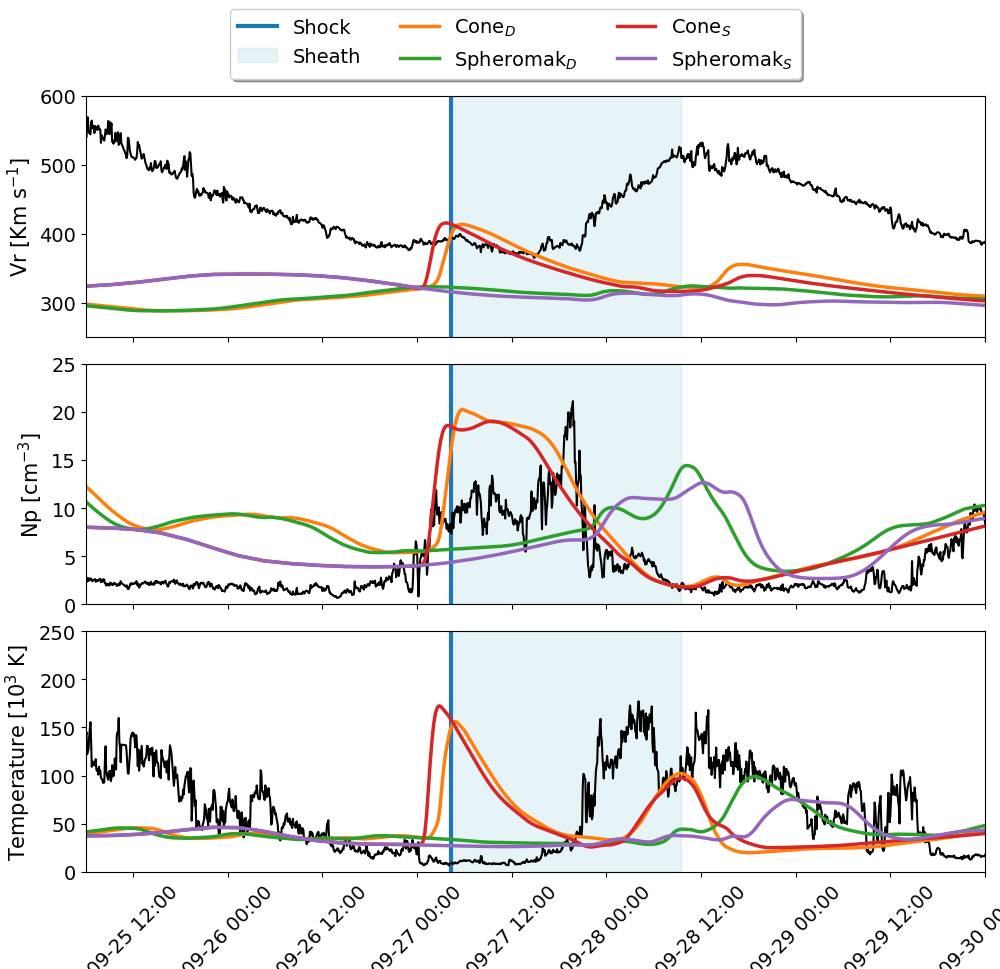}
  \caption{Time series of the magnetic (left) and plasma (right) variables at ACE. The figure is given in the same format as Fig.~\ref{fig:cone_spheromak_divv_v_trp_Bepi}. } \label{fig:L1_B_plasma}
\end{figure*}

\subsection{Parker Solar Probe} \label{appendix_psp}
The evolved CME models upon the arrival at PSP are shown in Figure~\ref{fig:cone_spheromak_divv_v_trp_PSP}. The figure is arranged similarly to Figure~\ref{fig:cone_spheromak_divv_v_trp_Bepi}. The domain is extended to $\pm 240$ R$_\odot$ and the cone CME is significantly wider than the spheromak CME model in these snapshots. Table~\ref{table:icarus_arrival_psp} shows that the cone and spheromak CMEs arrive $\sim4$ and $\sim3$ hours later in the dynamic solar wind than in the steady solar wind, respectively. Here, the spheromak CME arrives slightly earlier than the cone CME in the dynamic solar wind.
In contrast, in all previous cases, the spheromak model propagated in the dynamic solar wind arrived last. The observed arrival time is better reproduced when the two CME models are propagated in the steady solar wind because the CME models are significantly delayed when the dynamic solar wind is used.

Figure~\ref{fig:PSP_B_plasma} shows the time series at PSP for the magnetic field (left) and plasma (right) parameters. The figure describing the magnetic field components is arranged the same way as Figure~\ref {fig:Bepi_B_plasma}. The plasma variables shown are the velocity, proton number density and temperature from top to bottom. Since there was significant uncertainty in the data, we plotted the data from two instruments onboard PSP as part of the Solar Wind Electrons Alphas and Protons (SWEAP) suite, namely the Solar Probe Cup (SPC) and the Solar Probe ANalyzers – Ions (SPI). From all quantities, it is evident that the SPC results are in better agreement with the simulated results and observations at other spacecraft.

The arrival time for the cone CME, which propagated in the steady solar wind, coincides with the arrival time in the observations. The speed profile is modelled well in all simulations. The number density is consistently overestimated again. The spheromak CMEs better replicate the temperature profile. However, the sheath region is significantly smaller in the simulations compared to the observations. 
The magnetic field is modelled approximately well in the magnetic cloud compared to the poorly modelled CME sheath, as it arrives later in the simulations than in the observations and is compressed. In particular, the spheromak CME in the dynamic solar wind attempts to mimic the observed structures; however, the sheath portion is shorter. 
In the magnetic field components, the observed profiles are replicated crudely. Finally, the cone CME fails to reconstruct the magnetic cloud of the ejecta completely, as expected, since it lacks its intrinsic magnetic field configuration. 

\begin{table}[ht!]   
\centering  
\caption{The arrival times {for the observed CME shock and} for the cone and spheromak CME models in the steady ($S$) and dynamic ($D$) solar wind regimes at ACE.}   
\begin{tabular}{lc}         
\hline 
 & Arrival time (UT) \\ 
\hline  \hline    
    {Observed} & {2021-09-27T04:16:00} \\
    Cone$_D$ & 2021-09-27T01:38:00  \\ 
    Cone$_S$ & 2021-09-27T00:41:00 \\ 
    Spheromak$_D$ & 2021-09-27T21:55:00 \\
    Spheromak$_S$ & 2021-09-27T21:55:00\\ 
\hline                           
\end{tabular}
\label{table:icarus_arrival_L1}
\end{table}

\subsection{ACE} \label{appendix_ACE}
Figure~\ref{fig:cone_spheromak_divv_v_trp_L1} is arranged similarly to the same figures in the previous two subsections and shows the solar wind-CME configuration at the disturbance arrival time at L1. The location of L1 is indicated on the figure. We can see that the cone CME arrives earlier than the spheromak CME. The longitudinal expansion of the cone CME allows it to reach the spacecraft at the flank. In contrast, the spheromak CME barely extends to the L1 point. 

Figure~\ref{fig:L1_B_plasma} confirms these observations in the time series. The figure is arranged the same way as Figure~\ref{fig:PSP_B_plasma}. We do not have the region denoting the magnetic ejecta here, since it was not clearly distinguished in the observations. In plasma quantities, the cone CME can replicate the arrival time of the disturbance and the global density variation. The magnetic field, velocity, and temperature profiles are not well modelled. We note, however, that two high-speed streams interact at L1, which were not reconstructed well with the input boundary conditions generated with the GONG-WSA {coronal} model. The profiles in the simulations were therefore modelled poorly due to the absence of the CME-high speed stream interaction. 

Table~\ref{table:icarus_arrival_L1} shows the arrival times {of the CME in the observations and }of the cone and spheromak CMEs in the steady and dynamic solar wind regimes. Considering the low quality of the modelled data for the spheromak simulations, we cannot draw significant conclusions from this, as the minor disturbance arriving in the simulation is smeared out. This is confirmed by the arrival times. The spheromak CME model arrives nearly at the same time in the steady and dynamic regimes, and $\sim20-21$ hours later than the cone CME in the dynamic and steady regimes, respectively. 
The main limitation of the spheromak model is in successfully reconstructing flank encounters, since it does not succeed in building a sheath extended enough in longitude, {most likely because of lacking the CME legs}. Next, the signatures of a magnetic cloud are not present in the observations, and we use these to establish the longitudinal bounds of the magnetic cloud extent in that direction. {As a result, the disturbance arrived earlier at L1 when propagating the cone CME model and much later and fainter with the spheromak CME model, hinting that in reality the CME configuration was somewhere in between these two models.} Finally, although we do not see the magnetic ejecta signatures at L1, we can see the contribution of the CME sheath at L1, which makes the current CME event a five-spacecraft CME event.

\end{appendix}
\end{document}